\documentclass[reprint,aps,prd,nofootinbib,floatfix,amsmath,amssymb]{revtex4-1}

\usepackage{graphicx}
\usepackage{multirow}
\usepackage{amssymb}
\usepackage[usenames]{color}
\usepackage[toc,page]{appendix}
\usepackage{slashed}
\usepackage{bbding}
\usepackage{booktabs}

\usepackage[colorlinks=true,urlcolor=blue,linkcolor=blue,citecolor=blue]{hyperref}

\newcommand{\decayS}{$\ell\to \ell^{\prime}\ell^{\prime\prime}\bar{\ell}^{\prime\prime\prime}$}
\newcommand{\eq}[1]{\begin{align}#1\end{align}}

\makeatother

\begin{document}

\title{Effects of heavy Majorana neutrinos on lepton flavor violating processes}

\author{G. Hern\'andez-Tom\'e}
\affiliation{
CAFPE and Departamento de F\'isica Te\'orica y del Cosmos, 
Universidad de Granada, E–18071 Granada, Spain}
\author{J.~I. Illana}
\affiliation{
CAFPE and Departamento de F\'isica Te\'orica y del Cosmos, 
Universidad de Granada, E–18071 Granada, Spain}
\author{G. L\'opez Castro}
\affiliation{
Departamento de F\'isica, 
Centro de Investigaci\'on y de Estudios Avanzados del Instituto Polit\'ecnico Nacional,
Apdo.~Postal 14-740, 07000 M\'exico D.F., M\'exico}
\author{M. Masip}
\affiliation{
CAFPE and Departamento de F\'isica Te\'orica y del Cosmos, 
Universidad de Granada, E–18071 Granada, Spain}
\author{P. Roig}
\affiliation{
Departamento de F\'isica, 
Centro de Investigaci\'on y de Estudios Avanzados del Instituto Polit\'ecnico Nacional,
Apdo.~Postal 14-740, 07000 M\'exico D.F., M\'exico}

\date{\today}

\begin{abstract}
The observation of lepton flavor violating processes at colliders could be a clear signal of a non-minimal neutrino sector. We define a 5-parameter model with a pair of TeV fermion singlets and arbitrary mixings with the three active neutrino flavors. Then we analyze several flavor violating transitions ($\ell\to \ell^{\prime}\gamma,\,\ell^{\prime}\ell^{\prime\prime}\bar{\ell}^{\prime\prime\prime}$ or $\mu-e$ conversions in nuclei) and $Z\to\bar{\ell}\,\ell^{\prime}$ decays induced by the presence of heavy neutrinos. In particular, we calculate all the one-loop contributions to these processes and present their analytic expressions. We focus on the  genuine effects of the heavy Majorana masses, comparing the results in that case with the ones obtained when the two heavy neutrinos define a Dirac field. Finally, we use our results to update the bounds on the heavy-light mixings in the neutrino sector.
\end{abstract}

\maketitle


\section{Introduction}
\label{Intro}

In the original formulation of the standard model (SM) \cite{Glashow:1961tr, Weinberg:1967tq, Salam:1968rm}, the lepton flavor and the lepton number are accidentally conserved quantities due to the assumption of massless neutrinos. However, this framework must be extended to account for the well-established  evidence of neutrino oscillations \cite{Fukuda:1998mi, Ahmad:2001an, Ahmad:2002jz}, which implies non-zero masses and mixings for the active neutrinos. A possible minimal extension is the so-called $\nu$SM \cite{Mohapatra:1998rq}, which adds right-handed components (gauge singlets) for the three neutrino families and generates Dirac masses via Yukawa couplings with the Higgs doublet, just like for all the other fermions. In the $\nu$SM, the mixing in the leptonic sector is described by a $3\times 3$ unitary matrix called the PMNS matrix \cite{Pontecorvo:1957qd,Maki:1962mu}, analogous to the CKM matrix of the quark sector \cite{Cabibbo:1963yz,Kobayashi:1973fv}. Nevertheless, the $\nu$SM requires extremely tiny Yukawa couplings to explain the observed masses, which suggests that other mechanism may be at work. If, in addition to the Dirac mass terms ($m_D$) that combine them with the active neutrinos, the singlets have Majorana masses ($m_M$) that define a new scale, then the tiny neutrino masses appear naturally for a very large value of $m_M$ ({\it i.e.}, $m_M \gg m_D$). In this seesaw mechanism \cite{Minkowski:1977sc, GellMann:1980vs, Mohapatra:1979ia} the new mass terms break lepton number. The physical states after diagonalization of the mass matrix include light ($\nu$) and heavy ($N$) sectors of Majorana neutrinos with masses 
\eq{
m_\nu\approx m_D^2/m_M, \quad m_N \approx m_M \gg m_\nu.
}
Like in the $\nu$SM, in this model the mixings among the three active families may be large, as required from oscillation experiments, but the mixing with the heavy fields is of order 
\eq{
s_\nu\approx m_{D}/m_{M}\approx \sqrt{m_\nu/m_N}.\label{ss-relation}
}

Both in the $\nu$SM and this high scale (type I) seesaw model, the rate of lepton flavor violating (LFV) processes at colliders is suppressed by a factor of $(m_\nu/E)^2$, being $E$ the scale of the process. In the second scenario LFV can also be mediated by the neutrinos in the heavy sector, but the heavy-light mixing implies then a suppression of order of $(E/m_N)^2$, equally small. In particular, the LFV decays $\ell\to \ell^{\prime}\gamma$, \decayS\ and $Z\to \bar{\ell}\ell^{\prime}$, where $\ell,$ $\ell^{\prime},\dots$ denote the usual charged leptons ($\tau,$ $\mu$, $e$) will have a branching ratio below $10^{-50}$ \cite{Petcov:1976ff,Bilenky:1977du,Cheng:1985bj,Illana:1999ww, Illana:2000ic,Hernandez-Tome:2018fbq,Blackstone:2019njl}. It is then apparent that any experimental observation of flavor violation involving charged leptons (cLFV) would unambiguously imply the existence of new physics at the TeV scale in an extended neutrino sector \cite{Arganda:2004bz,Dinh:2012bp,Dinh:2013vya,Abada:2014cca,Arganda:2014dta, DeRomeri:2016gum,Lindner:2016bgg}.

Well-motivated variants of the two minimal models described above include the inverse seesaw \cite{Mohapatra:1986bd, Bernabeu:1987gr} or the linear seesaw \cite{Malinsky:2005bi}. These scenarios allow for arbitrary masses in the heavy neutrino sector and then unsuppressed heavy-light mixings, constrained only by the experimental limits. They are usually known as low-scale seesaw models, although the masses in both sectors are not necessarily correlated. They are justified by approximate symmetries or some ansatz on the neutrino mass matrix that relaxes the restriction in Eq.~(\ref{ss-relation}). This type of models may be adequate in scenarios like little Higgs (the heavy Majorana in seesaw models would introduce quadratic corrections to the Higgs mass \cite{delAguila:2005yi,delAguila:2017ugt,delAguila:2019mvp}), supersymmetry \cite{Abada:2014kba,Arganda:2015naa,Arganda:2015ija}, TeV gravity models (with the cutoff right above that scale) \cite{ArkaniHamed:1998rs,Randall:1999ee} or composite Higgs models \cite{Coito:2019wte}. In the next section we present a simple model that captures all the relevant effects that may appear in cLFV processes induced by the presence of heavy neutrinos. 

Another possibility in these scenarios that is interesting from the phenomenological 
point of view is to test the Dirac or Majorana nature of the neutrinos in the heavy sector through lepton number ($L$) violating processes with $\Delta L=2$. Apart from the longly explored neutrinoless double-beta decay \cite{Furry:1939qr,Zeldovich:1981da}, this has been undertaken in tau decays $\tau^-\to\ell^+M_1^-M_2^-$ ($M_1, M_2=\pi, \, K$ mesons) \cite{Ilakovac:1995km, Ilakovac:1995wc, Gribanov:2001vv}, meson decays  $M_1^+\to\ell_1^+\ell_2^+M_2^-$ and hyperon decays, like $\Sigma^-\to\Sigma^+e^-e^-$, $\Sigma^-\to p\mu^-\mu^-$ \cite{Littenberg:1991rd, Barbero:2002wm}, etc. All these studies are based on scenarios where the new sterile Majorana neutrinos have non-negligible mixings and some of them require masses low enough to be produced on-shell (resonant-enhancement).

\begin{table*}
\caption{Present limits and future sensitivities for the branching ratios or capture rates of several LFV processes. We denote $Z\to\ell_1\ell_2\equiv Z\to\bar\ell_1\ell_2+\ell_1\bar\ell_2$, and similarly for $h$ decays. For a more extensive list including hadronic modes see \cite{Calibbi:2017uvl, Amhis:2016xyh}. }\label{cLFV limits}
\begin{ruledtabular}
\begin{tabular}{cccccc}
Reaction & Present Limit 90\% C.L.  & Future Sensitivity & 
Reaction & Present Limit 90\% C.L.  & Future Sensitivity 
\\ \hline
$\mu\to e\gamma$ & $4.2 \times 10^{-13}$ \cite{Adam:2013mnn} & 
$6 \times 10^{-14}$ \cite{Baldini:2018nnn} & $\mu\to ee\bar{e}$ & 
$1.0 \times 10^{-12}$ \cite{Bellgardt:1987du} & $10^{-16}$ \cite{Blondel:2013ia}    
\\ \hline
$\mu-e$ (Au) & $7.0 \times 10^{-13}$ \cite{Bertl:2006up} & --- & 
$\mu-e$ (Ti) & $4.3 \times 10^{-12}$ \cite{Bertl:2006up} & 
$10^{-18}$ \cite{Alekou:2013eta}
\\ \hline
$\tau\to e\gamma$  & $3.3 \times 10^{-8}\,$ \cite{Aubert:2009ag} & 
$3\times 10^{-9}$ \cite{Kou:2018nap}  & 
$\tau\to \mu \gamma$ & $4.4 \times 10^{-8}\,$ \cite{Aubert:2009ag}  & 
$10^{-9}$ \cite{Kou:2018nap}
\\ \hline
$\tau\to ee\bar{e}$ & $2.7 \times 10^{-8}$ \cite{Tanabashi:2018oca}  &   & $\tau\to \mu\mu\bar{\mu}$ & $2.1 \times 10^{-8}\,$  \cite{Tanabashi:2018oca}  & 
\\ 
$\tau\to e\mu\bar{\mu}$ & $2.7 \times 10^{-8}\,$ \cite{Tanabashi:2018oca} & $(2-5)\times 10^{-10}$ \cite{Kou:2018nap} & 
$\tau\to \mu e\bar{e}$  & $1.8 \times 10^{-8}\,$ \cite{Tanabashi:2018oca} & $(2-5)\times 10^{-10}$ \cite{Kou:2018nap}
\\ 
$\tau\to ee\bar{\mu}$   & $1.5 \times 10^{-8}\,$ \cite{Tanabashi:2018oca} &  & $\tau\to \mu\mu\bar{e}$ & $1.7 \times 10^{-8}\,$ \cite{Tanabashi:2018oca} & 
\\ 
\hline\hline
Reaction & Present Limit 95\% C.L.  & Future Sensitivity & 
Reaction & Present Limit 95\% C.L.  & Future Sensitivity 
\\ \hline
$Z\to \mu e$ & $7.3 \times 10^{-7}$ \cite{Nehrkorn:2017fyt}  & 
$10^{-10}$ \cite{Dam:2018rfz} 
  & $h\to \mu e$ & $3.4 \times 10^{-4}$ \cite{Khachatryan:2016rke}    & --- 
\\
$Z\to \tau e$ & $9.8 \times 10^{-6}$ \cite{Akers:1995gz} & 
$10^{-9}$ \cite{Dam:2018rfz}  & $h\to \tau e$ & $6.2 \times 10^{-3}$ \cite{Sirunyan:2017xzt} & \multirow{2}{*}{$5\times 10^{-4}\,$\cite{Cerri:2018ypt}}    \\ 
$Z\to \tau\mu$  & $1.2 \times 10^{-5}$ \cite{Abreu:1996mj}  & 
$10^{-9}$ \cite{Dam:2018rfz}  & $h\to \tau\mu $ & $2.5 \times 10^{-3}$ \cite{Sirunyan:2017xzt}  & 
\\
\end{tabular}
\end{ruledtabular}
\end{table*}

Currently there is no evidence for cLFV, but intense experimental efforts have provided strong limits on an extensive list of processes; some of them are reported in Table~\ref{cLFV limits}. The sensitivity to these transitions will be considerably improved in near-future experiments. The MEG-II and Mu3e experiments will reach branching ratios of order $6 \times 10^{-14}$ \cite{Baldini:2018nnn} and $10^{-16}$ \cite{Blondel:2013ia} for $\mu \to e \gamma$ and  $\mu \to e e\bar{e}$, respectively, whereas the expected bounds from PRISM and COMET will be near $10^{-18}$ \cite{Alekou:2013eta} for $\mu-e$ (Ti) conversion and $10^{-17}$ \cite{Kuno:2013mha} for $\mu-e $ (Al). For the third family, the  bounds on the $\tau\to\ell^{\prime}\gamma$ and $\tau\to\ell^{\prime}\ell^{\prime\prime}\bar{\ell^{\prime\prime\prime}}$ branching ratios could be improved by two orders of magnitude at Belle-II when the experiment achieves its maximum luminosity \cite{Hays:2017ekz,Kou:2018nap}. LHCb has already set a stringent limit (competitive with the present ones at Belle) of $4.6\times 10^{-8}$ \cite{Aaij:2014azz} on the $\tau\to\mu\mu\bar{\mu}$ process. In its high-luminosity phase the LHC is expected to improve this bound by one order of magnitude. Additionally, the possibility of running at the $Z$ pole in the electron-positron version of a Future Circular Collider (FCC-ee) \cite{Dam:2018rfz,Blondel:2019yqr} or in the Circular Electron Positron Collider (CEPC) \cite{CEPCStudyGroup:2018ghi} would improve the current limits on $Z\to\bar{\ell}\ell^{\prime}$ by about four orders of magnitude. Finally, the expected sensitivity of the HL-LHC ($3000\, {\rm fb}^{-1}$) will be around $5\times10^{-4}$ for both the $h \to e \tau$ and $h \to \mu \tau$ branching fractions \cite{Cerri:2018ypt}.

In this work we focus on the most phenomenologically relevant cLFV observables in the framework of low-scale seesaw scenarios. In Section \ref{TF} we introduce a model for the mixings of the active neutrinos with two singlet fermions defining Majorana fields. The mass splitting between these heavy fields parametrizes the breaking of lepton number; when the splitting vanishes the heavy sector reduces to a single Dirac neutrino. In Section \ref{cLFV-amplitudes} we provide detailed expressions for the amplitudes and decay rates of the processes under consideration. In Section \ref{NA} we use these observables to derive constraints on the heavy-light mixing angles as a function of the masses of the two heavy states. Our conclusions are given in Section \ref{Conclusions}. 


\section{A model for the heavy-light neutrino mixing}
\label{TF}

As mentioned before, in the usual type-I seesaw model with one Majorana singlet per family the heavy-light mixings are correlated with the neutrino masses: to obtain $m_\nu<1$~eV with $m_D\approx 1$ GeV we need $m_M>10^9$ GeV, and this implies negligible heavy-light mixings, $s_\nu<10^{-9}$. As it is well known by the practitioners, however, this is no longer the case when the singlet fermions are introduced in pairs (see \cite{Bolton:2019pcu} for a recent review). 
In particular, all the heavy-light mixing effects can be captured by considering a model with just one extra pair. Let us see how this works. 

We take two bi-spinors $N$ and $N^c$ of left-handed chirality ({\it undotted}), sterile
and with opposite lepton number, and define the four-spinors
\eq{
N_L = \begin{pmatrix} N \\ 0 \end{pmatrix}\,;\quad
N_R = \begin{pmatrix} 0 \\ \bar N^c \end{pmatrix}\,;\quad
\nu_{Li} = \begin{pmatrix} \nu_i \\ 0 \end{pmatrix}\,,
}
where $\nu_{i=e,\mu,\tau}$ are the SM neutrinos. After the breaking of the electroweak symmetry the SM charged leptons get masses through Yukawa interactions with the Higgs field; the left-handed mass eigenstates are obtained after a unitary transformation,
\eq{
\ell_{Li} \leadsto \sum_{j=1}^{3}U_{ij}^{\ell}\ell_{Lj},
}
that we also perform in the space of the three active neutrinos $\nu_{Li}$. Then we assume that in this basis the 5 Majorana fields $\chi_i=\chi_{Li}+\chi_{Li}^c$ with  $\chi_L\equiv(\nu_{L1},\nu_{L2},\nu_{L3},N_L,N^c_R)$ have mass terms
\eq{
{\cal L}_M &= -\frac{1}{2}\overline{\chi^c_{Li}}\,{\cal M}\,\chi_{Lj} + {\rm h.c.}, 
}
with
\eq{
{\cal M} = \begin{pmatrix}
        0   & 0   & 0   & 0 & m_1 \\
        0   & 0   & 0   & 0 & m_2 \\
        0   & 0   & 0   & 0 & m_3 \\
        0   & 0   & 0   & 0 & M   \\
        m_1 & m_2 & m_3 & M & \mu 
\end{pmatrix}.
\label{numatrix}
}
Notice that we have ordered the fields according to their lepton number (positive for the
first four neutrinos), and that the Majorana mass $\mu$ corresponds to the neutrino with negative
lepton number (the fifth one).
The mass eigenstates are obtained diagonalizing this symmetric matrix by an orthogonal transformation and applying a field redefinition ($\chi_{L4}\to i\chi_{L4}$) to guarantee real and positive mass eigenvalues. Three eigenvalues are zero ($m_{\chi_{1,2,3}}=0$) and the other two are 
\eq{
m_{\chi_4}&=
{1\over 2} \left(\sqrt{4\left( m_1^2+m_2^2+m_3^2+M^2\right) + \mu^2} - \mu \right) , \\
m_{\chi_5}&=
{1\over 2} \left(\sqrt{4\left( m_1^2+m_2^2+m_3^2+M^2\right) + \mu^2} + \mu \right) .
}
Defining $m\equiv \sqrt{m_1^2+m_2^2+m_3^2}$ and $M'\equiv \sqrt{m^2+M^2}$, the mass eigenstates are given by the replacement
\eq{
\chi_{Li} \leadsto U_{ij}^\nu \chi_{Lj},
}
where the mixing matrix reads
\begin{widetext}
\eq{
U^\nu=
\begin{pmatrix} 
-{m_2\over \sqrt{m_1^2+m_2^2}} & 
-{m_1m_3\over m\sqrt{m_1^2+m_2^2}} & 
-{m_1M\over mM'} &
-i{m_1m_{\chi_5}\over M'\sqrt{m_{\chi_5}^2+M'^2}} & 
{m_1\over \sqrt{m_{\chi_5}^2+M'^2}}
\cr 
{m_1\over \sqrt{m_1^2+m_2^2}} & 
-{m_2m_3\over m\sqrt{m_1^2+m_2^2}} & 
-{m_2M\over mM'} & 
-i{m_2m_{\chi_5}\over M'\sqrt{m_{\chi_5}^2+M'^2}} &
{m_2\over \sqrt{m_{\chi_5}^2+M'^2}}
\cr
0 & 
{\sqrt{m_1^2+m_2^2}\over m} & 
-{m_3M\over mM'} & 
-i{m_3m_{\chi_5}\over M'\sqrt{m_{\chi_5}^2+M'^2}} &
{m_3\over \sqrt{m_{\chi_5}^2+M'^2}}
\cr
0 &
0 & 
{m\over M'} & 
-i{Mm_{\chi_5}\over M'\sqrt{m_{\chi_5}^2+M'^2}} & 
{M'\over \sqrt{m_{\chi_5}^2+M'^2}} 
\cr
0 &
0 &
0 &
i{M'\over \sqrt{m_{\chi_5}^2+M'^2}} &
{m_{\chi_5}\over \sqrt{m_{\chi_5}^2+M'^2}} 
\end{pmatrix} .
\label{massmatrix}
}
\end{widetext}

Several comments are in order: 

\begin{itemize}

\item
$\mu$ is the only mass parameter breaking lepton number.
This parameter defines the mass splitting of 
the two heavy Majorana neutrinos: when $\mu=0$ both states form 
a heavy Dirac neutrino singlet of mass $M'$.

\item
The two heavy neutrinos ($N_{1,2}\equiv\chi_{4,5}$) have components of order $m_i/M$ along the corresponding active neutrinos, where $m_i=Y_{\nu_i}v/\sqrt{2}$ are Dirac mass terms coming from Yukawa couplings with the SM Higgs doublet. If the heavy fields have TeV masses, the heavy-light mixings can be as large as $\sim 0.1$ for couplings of order one.

\item
The three (mostly) active neutrinos ($\chi_{1,2,3}$), to be identified with the light neutrinos ($\nu_{1,2,3}$) observed so far, are exactly massless. A deformation of the pattern
in Eq.~(\ref{numatrix})
with a Majorana mass $\mu'$ for $N_L$ (like in inverse seesaw models) would imply that 
one of these neutrinos gets a mass $m_\nu\approx \mu' (m/M)^2$. Since 
$m_\nu< 1$ eV, however, the new term $\mu'$ must be small and has no effect on flavor physics
(it does not change the heavy-light mixings). An analogous argument applies 
to possible Dirac mass terms $m'_i$ in the fourth row/column.

\item
The pattern that we propose must be understood as approximate: these 5 entries  are the 
dominant mass terms, and any deformation must respect that the third neutrino 
gets a mass below 1 eV ({\it i.e.}, it must be much smaller than these 5 terms).
Notice also that no symmetry protects the entries assumed to be zero, 
and that loop corrections will actually introduce contributions to all of them
(see \cite{Bolton:2019pcu}). 
The pattern must then be established at the loop level where we work, which may require
the addition of tree level terms canceling radiative corrections. In particular, large 
values of  $\mu$ would induce 1-loop values of $\mu'$ that must be canceled 
to obtain the proposed pattern. This fine tuning disappears for $\mu\to 0$, when
the resulting pattern is justified by the conservation of lepton number.

\item
The generation of small masses and light-light mixings for the three 
active neutrinos would require
the addition of extra singlets. This could be accommodated with the usual mechanisms 
(in $\nu$SM or Type I seesaw models) or through the deformations described above (in inverse 
seesaw models). In any case, it will not introduce sizeable heavy-light mixings. 

\end{itemize}

We will trade the five arbitrary mass parameters in Eq.~(\ref{massmatrix}) for the masses of the two heavy neutrinos and three heavy-light mixings, 
\eq{
m_{N_1} \equiv m_{\chi_4}, \quad
m_{N_2} \equiv m_{\chi_5}, \quad 
s_{\nu_i}\equiv {m_i\over \sqrt{m_{N_1} m_{N_2}}}.
}

The $5\times 5$ matrix $U_{ij}^\nu$ above will introduce tree-level charged and neutral currents involving neutrinos:
\eq{
\mathcal{L}_{W^{\pm}}&=-\frac{g}{\sqrt{2}}W^{-}_{\mu}\sum_{i=1}^{3} \sum_{j=1}^{5}B_{ij}\,\bar{\ell}_{i}\gamma^{\mu}P_L\chi_{j}+{\rm h.c.},\label{cc}
\\
\mathcal{L}_{Z}&=-\frac{g}{4c_W}Z_{\mu} \sum_{i,j=1}^{5}\bar{\chi}_{i}\gamma^{\mu}\left(C_{ij}P_L -C_{ij}^{*}P_R \right)\chi_{j},\label{nc} 
\\
\mathcal{L}_{G^{\pm}}&=-\frac{g}{\sqrt{2}M_W}G^{-}\sum_{i=1}^{3} \sum_{j=1}^{5}B_{i j} \nonumber\\
&\qquad\qquad\quad \times\bar{\ell}_{i}\left(m_{\ell_{i}}P_L-m_{\chi_j} P_R\right)\chi_{j}+{\rm h.c.},\label{Gc}
}
where $G^\pm$ is the charged would-be-Goldstone field, $g$ is the weak coupling constant, $c_W=\cos\theta_W$ and $P_{L,R}=\frac{1}{2}(1\mp \gamma_5)$ are the left and right-handed projectors, respectively. Notice that in Eq.~(\ref{nc}), the neutral current induced by the Majorana states involves couplings of different flavors  with both left and right-handed components.\footnote{%
For the case of heavy left-handed neutrinos being sequential Dirac (active) neutrinos, replace $B_{ij}\to U^\nu_{ij}$, $C_{ij}\to\delta_{ij}$, $C_{ij}^{*}\to 0$.}
The dimension of the rectangular $B$ mixing matrix is $3\times 5$, whereas $C$ is a 
$5\times 5$ matrix,
\eq{
B_{ij}=\sum_{k=1}^{3}\delta_{ik} U^{\nu}_{kj}\,,\quad C_{ij}=\sum_{k=1}^{3}{(U_{ki}^{\nu})}^{*}U_{kj}^{\nu}.
}
One can see that the elements of these matrices involving heavy neutrinos can be expressed in terms of heavy-light mixings and the squared mass ratio $r=m^2_{N_2}/m^2_{N_1}$ as
\eq{
B_{k N_{1}}=-\frac{i\,r^{\frac{1}{4}}}{\sqrt{1+r^{\frac{1}{2}}}}s_{\nu_k}, \quad 
B_{k N_{2}}=\frac{1}{\sqrt{1+r^{\frac{1}{2}}}}s_{\nu_k},
\label{Bsfortwo}
}
\eq{
C_{N_{1}N_{1}}&=\frac{r^{\frac{1}{2}}}{1+r^{\frac{1}{2}}}\sum_{k=1}^{3}s_{\nu_k}^{2}, \quad 
C_{N_{2}N_{2}}=\frac{1}{1+r^{\frac{1}{2}}}\sum_{k=1}^{3}s_{\nu_k}^{2},
\nonumber\\
&C_{N_{1}N_{2}}=-C_{N_{2}N_{1}}=\frac{i\,r^{\frac{1}{4}}}{1+r^{\frac{1}{2}}}\sum_{k=1}^{3}s_{\nu_k}^2 .\label{Csfortwo}
}
These are the same as in \cite{Ilakovac:1994kj, Illana:2000ic} up to an irrelevant global phase for $B$. In addition, the matrices $B$ and $C$ satisfy some identities that are essential to keep the renormalizability of the model:
\eq{
\sum_{k=1}^{5}B_{ik}B_{jk}^{*}&=\delta_{ij}, \quad  
\sum_{k=1}^{3}B_{k i}^{*}B_{k j}=\sum_{k=1}^{5}C_{ik}C_{jk}^{*}=C_{ij},
\nonumber\\
&\sum_{k=1}^{5}B_{ik}C_{kj}=B_{ij}, \label{identities1}
\\
\sum_{k=1}^{5}m_{\chi_k}C_{ik}C_{jk}&=\sum_{k=1}^{5}m_{\chi_k}B_{i k}C_{kj}^{*}=\sum_{k=1}^{5}m_{\chi_k}B_{i k}B_{j k}=0. \label{identities2}
}


\section{LFV Processes}
\label{cLFV-amplitudes}

We now present the amplitudes and decays widths or transition rates for the LFV processes $\ell\to \ell^{\prime}\gamma$, $Z\to \bar{\ell}\ell^{\prime}$, $\ell\to \ell^{\prime}\ell^{\prime\prime}\bar{\ell}^{\prime\prime\prime}$ and $\mu-e$ conversion in nuclei. All of them involve the effective interaction of a neutral vector boson with a pair of on-shell fermions, $V\ell \ell^{\prime}$ ($V=\gamma,\, Z$), through a loop with Majorana neutrinos. Since the $W$ couples only to left-handed fields, the effective  $V\ell \ell^{\prime}$ vertices ($\ell\ne\ell'$) can be written in terms of the following form factors:
\eq{
i\Gamma_{\mu}^{\gamma}(q^2)&=ie\big[F_L^{\gamma}(q^2)\gamma_{\mu}P_L+i2 F_M^{\gamma}(q^2)P_R\sigma_{\mu\nu}q^{\nu}
\big], \label{Effective-gammaffs}
\\
i\Gamma_{\mu}^{Z}(q^2)&=ie\big[F_L^{Z}(q^2)\gamma_{\mu}P_L\big],\label{Effective-Zffs}
}
where $q$ is the momentum of the $V$ boson. Actually, the most general Lorentz structure for on-shell fermions contains two additional (anapole) form factors, $F_S$ and $F_P$. However, they do not contribute when the $V$ boson is on-shell, due to the transversality condition $q_{\mu}\epsilon^{\mu}=0$. The same happens for an off-shell $V$ boson when the masses of the external fermions can be neglected \cite{Hollik:1998vz}. On the other hand, the dipole form factors (chirality flipping) are proportional to the external lepton masses. 

\begin{figure}
   \includegraphics[scale=0.55]{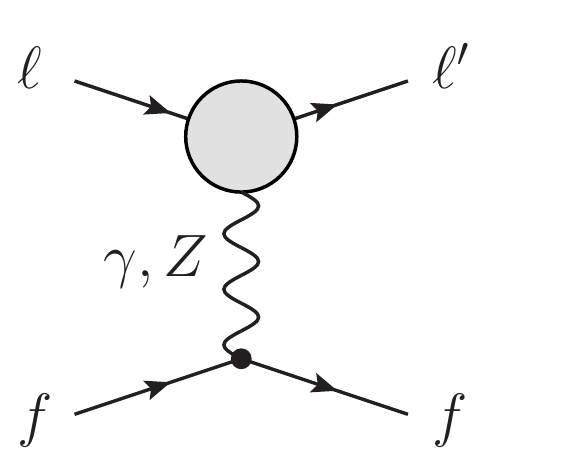} \quad
   \includegraphics[scale=0.55]{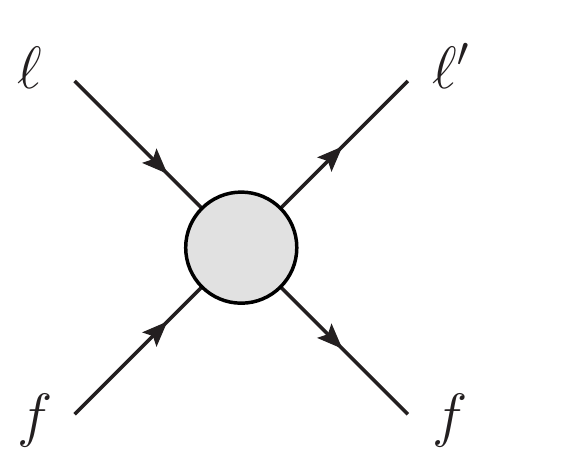} 
\caption{Generic penguin and box diagrams contributing to $\ell\to\ell'\ell''\bar{\ell}'''$ ($f=\ell''=\ell'''$) and $\mu-e$ conversion in nuclei ($f=u,d$).}
       \label{pen-and-box-F}          
\end{figure}

In the limit $q^2\to 0$, appropriate for $\ell\to\ell^{\prime}\gamma$ and for the penguin contributions to $\ell\to \ell^{\prime}\ell^{\prime\prime}\bar{\ell}^{\prime\prime\prime}$ and $\mu-e$ conversion (Fig.~\ref{pen-and-box-F}), we may write: 
\eq{
F_L^{\gamma}(q^2)\equiv q^2 A_{1L}, \quad F_M^{\gamma}(q^2)\simeq F_M^{\gamma}(0)\equiv \frac{m_{\ell}}{2}A_{2R}.
\label{FLA1L}
}
The vector form factor $F_L^{\gamma}$ for an on-shell photon vanishes by current conservation due to the electromagnetic gauge invariance, and hence only the dipole form factor $F_M^\gamma$ contributes to $\ell\to\ell^{\prime}\gamma$. Then the amplitude reads:  
\eq{
\mathcal{M}(\ell\to\ell^{\prime}\gamma)=ie m_{\ell}A_{2R}\bar{u}(p_{\ell^\prime})P_R\sigma^{\mu\nu}q_{\nu}u(p_{\ell})\epsilon_{\mu}^{\gamma}(q),
}
where $\epsilon_\mu^\gamma$ is the photon polarization vector and we have neglected the mass of the lighter lepton. The partial decay width is given by
\eq{
\Gamma(\ell\to \ell^{\prime}\gamma)=\alpha m_{\ell}^3|F_M^{\gamma}(0)|^2.
}

The $Z\to\bar{\ell}\ell^{\prime}$ decay proceeds through the $Z\ell\ell'$ vertex with $q^2=M_Z^2$. Here we can take both external leptons as massless and ignore the corresponding dipole form factor $F_M^Z$, hence omitted in (\ref{Effective-Zffs}). The amplitude is then given by
\eq{
\mathcal{M}(Z\to\bar{\ell}\ell^{\prime})=e F_{L}^{Z}(M_Z^2)\bar{u}(p_{\ell^\prime})\gamma^{\mu}P_L v(p_{\ell})\epsilon_{\mu}^{Z}(q),
}
where $\epsilon_\mu^Z$ is the $Z$ polarization vector and the partial decay width is 
\eq{
\Gamma(Z\to\bar{\ell}\ell^{\prime})=\frac{\alpha}{3}M_Z|F_L^Z(M_Z^2)|^2.
}

\begin{table}
\caption{Possible LFV \decayS\ channels.} \label{channels}
\begin{ruledtabular}
\begin{tabular}{cllll}
Type & Flavors & \multicolumn{3}{l}{\decayS}
\\ \hline
1 & $\ell\ne\ell^\prime=\ell^{\prime\prime}=\ell^{\prime\prime\prime}$ &
$\mu\to ee\bar{e}$ & $\tau\to ee\bar{e}$ & $\tau\to\mu\mu\bar{\mu}$ 
\\
2 & $\ell\ne\ell^\prime\ne\ell^{\prime\prime}=\ell^{\prime\prime\prime}$ && $\tau\to e\mu\bar{\mu}$ & $\tau\to\mu e\bar{e}$ 
\\
3 & $\ell\ne\ell^\prime=\ell^{\prime\prime}\ne\ell^{\prime\prime\prime}$ && $\tau\to ee\bar{\mu}$ & $\tau\to\mu\mu\bar{e}$ 
\\
\end{tabular} 
\end{ruledtabular}
\end{table}

Regarding \decayS, we distinguish the three types of decays in Table \ref{channels}.   Apart from the photon-penguin and $Z$-penguin diagrams containing the effective $V\ell\ell'$ vertices, these decays involve box diagrams (Fig. \ref{pen-and-box-F}):
\eq{
\mathcal{M}(\ell\to\ell^{\prime}\ell^{\prime\prime}\bar{\ell}^{\prime\prime\prime})=\mathcal{M}_{\gamma}+\mathcal{M}_{Z}+\mathcal{M}_B,
\label{ampli3lep}
}
where
\eq{
\mathcal{M}_{\gamma}&=\frac{e^2}{q^2} \bar{u}({p_{\ell^{\prime}}})\left(q^2 A_{1L}\gamma_{\mu}P_L+i m_\ell A_{2R}P_R\sigma_{\mu\nu}q^{\nu}\right)  u({p_{\ell}})\nonumber
\\
&\times \bar{u}({p_{\ell^{\prime\prime}}})\gamma^{\mu} v({p_{\ell^{\prime\prime\prime}}})-\left(\ell^{\prime} \leftrightarrow \ell^{\prime\prime} \right),\label{photon-contributions}
\\
\mathcal{M}_{Z}&=-\frac{e^2}{M_Z^2}F_L^{Z}(0) \bar{u}({p_{\ell^{\prime}}})\gamma_{\mu}P_L u({p_{\ell}})\nonumber
\\
&\times \bar{u}({p_{\ell^{\prime\prime}}})\gamma^{\mu}(g^{Z}_{L}P_{L}+g^{Z}_{R}P_{R}) v({p_{\ell^{\prime\prime\prime}}})-\left(\ell^{\prime} \leftrightarrow \ell^{\prime\prime} \right),\label{Z-contributions}
\\
\mathcal{M}_B&=e^2 F_{B} \bar{u}({p_{\ell^{\prime}}})\gamma_{\mu}P_L  u({p_{\ell}})\;\bar{u}({p_{\ell^{\prime\prime}}})\gamma^{\mu}P_L v({p_{\ell^{\prime\prime\prime}}}),\label{Box-contributions}
}
with $g_{L,R}^Z$ the charged lepton couplings to the $Z$ boson in units of $e$,
\eq{
g_L^Z=\frac{1}{2s_Wc_W}(-1+2s^2_W),\quad g_R^Z=\frac{s_W}{c_W},
\label{Zllcoup}
}
and the box diagrams are evaluated in the limit of zero external momenta. Channels of type 3 receive only box contributions, as they require two flavor-changing vertices. We have written the vector form factor $F_L^\gamma$ in terms of $A_{1L}$ (\ref{FLA1L}) to emphasize that the photon propagator cancels the $q^2$ prefactor. However, the dipole form factor $F_M^\gamma$, written here in terms of $A_{2R}$ (\ref{FLA1L}), will introduce a logarithmic dependence with the external lepton masses (they cannot be neglected) after phase-space integration of the squared amplitude. Notice that crossed diagrams with $\ell'$ and $\ell''$ exchanged must be added, except for channels of type 2. In the box amplitude, the form factor $F_B$ includes the crossed contribution thanks to a Fierz identity (see Eq.~\ref{A4} of Appendix~\ref{identities}). The expressions for the partial decay widths of \decayS\ as a function of $A_{1L}$, $A_{2R}$, $F_L^Z(0)$ and $F_B$ are given in Appendix~\ref{3lep}.

The  $\mu-e$ conversion in nuclei follows from similar diagrams as $\ell\to\ell'\ell''\bar{\ell}'''$ replacing the last two leptons by a quark $q=u$ or $d$ (Fig.~\ref{pen-and-box-F}). It involves the same photon-penguin and $Z$-penguin and a couple of new box form factors, $F_{B_u}$, $F_{B_d}$:
\eq{
{\cal M}_B^q= e^2 F_{B_q} 
\bar{u}(p_\mu)\gamma_\mu P_L u(p_q)\;
\bar{u}(p_e)\gamma^\mu P_L u(p'_q).
\label{Box-quarks}
}
The expressions for the $\mu-e$ conversion rate in nuclei as a function of $A_{1L}$, $A_{2R}$, $F_L^Z(0)$, $F_{B_u}$ and $F_{B_d}$ are given in Appendix~\ref{muecon}.

\begin{figure}
   \includegraphics[scale=0.55]{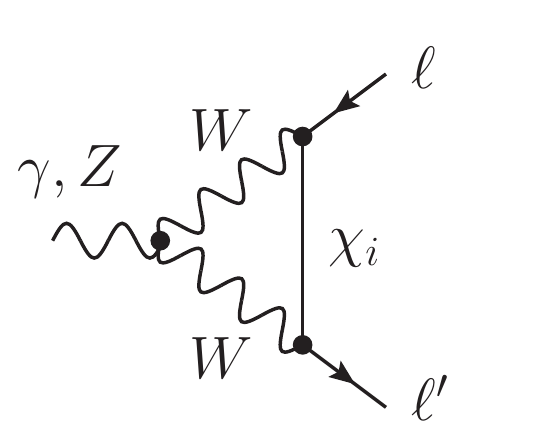} \quad
   \includegraphics[scale=0.55]{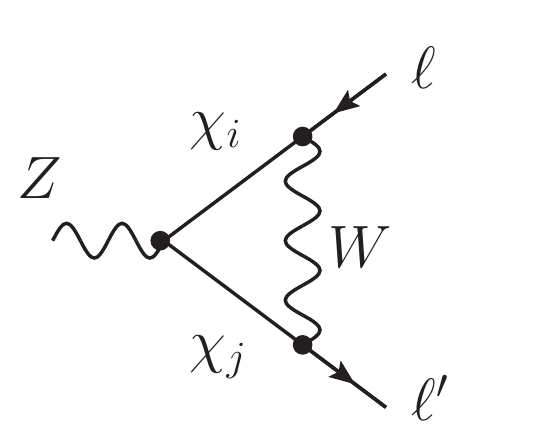} \quad
   \includegraphics[scale=0.55]{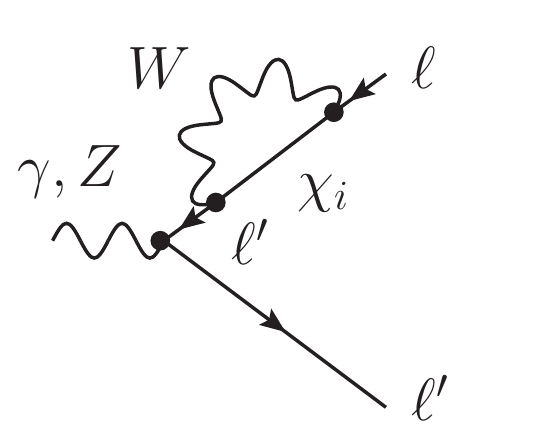} \quad
   \includegraphics[scale=0.55]{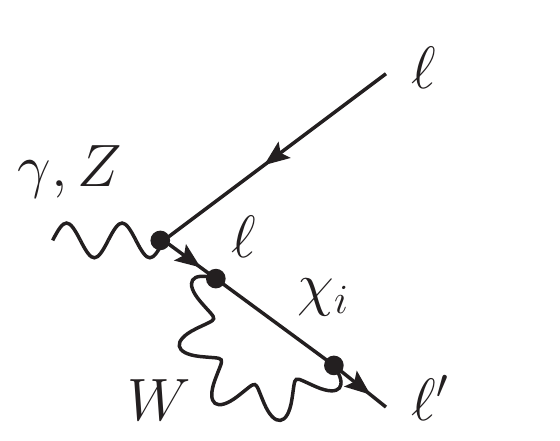}
\caption{One-loop diagrams contributing to the $V\ell\ell'$ vertex. We have omitted here and elsewhere diagrams with would-be-Goldstone fields, needed in the Feynman-'t Hooft gauge.}
       \label{Vll-Fig}          
\end{figure}

We have calculated in our model the one-loop contributions to the form factors introduced above, in the Feynman-'t Hooft gauge and using dimensional regularization. The effective  LFV $V\ell\ell'$ vertex is obtained from the diagrams of Fig.~\ref{Vll-Fig} supplemented by similar ones with the $W^\pm$ fields replaced by the would-be-Goldstone fields $G^\pm$. The resulting photon form factors in the low $q^2$ limit are:
\eq{
  F_{L}^{\gamma}(q^2)&= \frac{\alpha_W}{8\pi M_W^2}\sum_i^{5} B^{*}_{\ell i}B_{\ell^{\prime}i} f_L^{\gamma}(x_i;q^2),\label{FL}\\
  F_{M}^{\gamma}(0)&=\frac{\alpha_W}{8 \pi M_W^2}\frac{m_\ell}{2}\sum_i^{5} B^{*}_{\ell i}B_{\ell^{\prime}i} f_M^{\gamma}(x_i), \label{FM}
}
where $\alpha_W=\alpha/s_W^2$, $x_i\equiv m_{\chi_i}^2/M_W^2$ and
\eq{
   f_L^{\gamma}(x;q^2)&=\bigg[\frac{x^2
   \left(x^2-10 x+12\right) \ln x}{6
   (x-1)^4}\nonumber\\&+\frac{ \left(7 x^3-x^2-12
   x\right)}{12 (x-1)^3}-\frac{5}{9}\bigg]q^2+2 M_W^2\Delta_{\epsilon},
\\
f_M^{\gamma}(x)&= \frac{3 x^3 \ln x}{2 (x-1)^4}-\frac{2
 x^3+5 x^2-x}{4 (x-1)^3}+\frac{5}{6}.
}
The term  $\Delta_{\epsilon}=1/\epsilon-\gamma_E+\ln4\pi+\ln(\mu^2/M_W^2)$ regulates the ultraviolet divergence in $4-\epsilon$ dimensions and cancels in (\ref{FL}) for $\ell\ne\ell'$ due to the properties of $B$ (\ref{identities1}). As expected, $F_L^\gamma(0)$ is zero.

From (\ref{FM}) one may derive the contribution $\delta a$ of heavy neutrinos to the muon dipole moment anomaly, $(g-2)/2$. Subtracting that of light (massless) neutrinos, it reads:
\eq{
\delta a = \frac{\alpha_W}{4\pi}\frac{m^2_\mu}{M_W^2}
\sum_{i=4}^5 |B_{\mu i}|^2
\left[\frac{3 x_i^3 \ln x_i}{2 (x_i-1)^4}-\frac{2 x_i^3+5 x_i^2-x_i}{4 (x_i-1)^3}\right]
}
that is negative, enhancing the disagreement with the current experimental measurement \cite{Tanabashi:2018oca}, but anyway negligible because the prefactor is $\sim4\times 10^{-11}$, $|B_{\mu N_i}|^2\le s_{\nu_\mu}^2\lesssim 10^{-3}$ and the absolute value of the remaining function is smaller than 0.5. 

Regarding the $F_L^Z$ form factor of the effective $Z\ell\ell'$ vertex we find
\eq{
F_{L}^{Z}(q^2)&=\frac{\alpha_W}{8\pi s_W c_W}
\sum_{i,j}^{5} B^{*}_{\ell i}B_{\ell^{\prime}j}
\big[\delta_{ij}F(x_i;q^2) \nonumber\\
&+C^*_{ij}G(x_i,x_j;q^2)+C_{ij}\sqrt{x_{i}x_{j}}H(x_i,x_j;q^2)\big],
\label{FLZ}
}
where
\eq{
F(x;q^2)&= 2c_W^2\left[q^2\left(\overline{C}_1+\overline{C}_2+\overline{C}_{12} \right)-6\overline{C}_{00}+1 \right] \nonumber\\
&-(1-2s_W^2)x\overline{C}_{00}-2s_W^2xM_W^2\overline{C}_0\nonumber\\&+\frac{1}{2}\left(1-2c_W^2 \right)\left[(2+x)\bar{B}_{1}+1 \right],\\G(x, y;q^2)&=-q^2\left(C_{0}+C_{1}+C_{2}+C_{12}\right)+2C_{00}-1
\nonumber\\
&-\frac{1}{2}x yM_W^2C_{0} ,\\
H(x, y;q^2)&=q^2C_{0}+\frac{1}{2}q^2 C_{12}-C_{00}+\frac{1}{4},
}
in full agreement with \cite{Illana:2000ic}. Here we have used the following shorthand notation for the standard Passarino-Veltman loop functions \cite{Passarino:1978jh},
\eq{
C(x,y)_{\{ 0, 1, 2, 12 \}}&=C_{\{ 0, 1, 2, 12  \}}(0,q^2,0;M_W^2, x M_W^2, y M_W^2),\\
\overline{C}_{\{0,1,2,12\}}(x)&=C_{\{0,1,2,12 \}}(0,q^2,0;x M_W^2,  M_W^2,  M_W^2),\\
\bar{B}_1(x)&=B_1(0; x M_W^2, M_W^2),
}
defined with the same conventions as the computer packages {\tt LoopTools} \cite{Hahn:1998yk} and {\tt Collier} \cite{Denner:2016kdg}, that we have employed for numerical evaluations. Analytic expressions for these functions in the low $q^2$ limit, appropriate for the $Z$-penguin contribution to \decayS, can be found in \cite{Ilakovac:1994kj} and have been cross-checked with the help of {\tt Package-X} \cite{Patel:2016fam}. They are:
\eq{
F(x;0)&=\frac{5 x^2 \ln x}{2(x-1)^2}\!-\!\frac{5 x}{2(x-1)}+\frac{1}{4}
  \!-\!\left(\frac{5}{2}\!-\!2s_W^2\right)\Delta_{\epsilon},\label{FZS}
\\
G(x,y;0)&=\frac{1}{2(x-y)}\left[\frac{(y-1)x^2\ln x}{(x-1)}-\frac{(x-1)y^2\ln y}{(y-1)}\right]\nonumber\\
&+\frac{1}{2}\left( \Delta_{\epsilon}-\frac{1}{2}\right),\label{GZS}
\\
H(x,y;0)&=\frac{1}{4(x-y)} \left[\frac{x(x-4)\ln x}{x-1}-\frac{y(y-4)\ln y}{y-1}\right]\nonumber\\
&-\frac{1}{4}\left(\Delta_{\epsilon}+\frac{1}{2}\right).
}
The ultraviolet divergences cancel in (\ref{FLZ}) using the properties of the mixing matrices (\ref{identities1}) and (\ref{identities2}).

\begin{figure}
\begin{center}
   \includegraphics[scale=0.55]{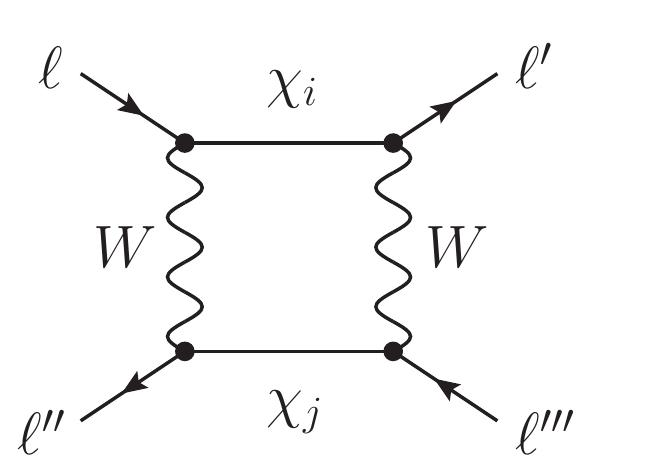} \quad
   \includegraphics[scale=0.55]{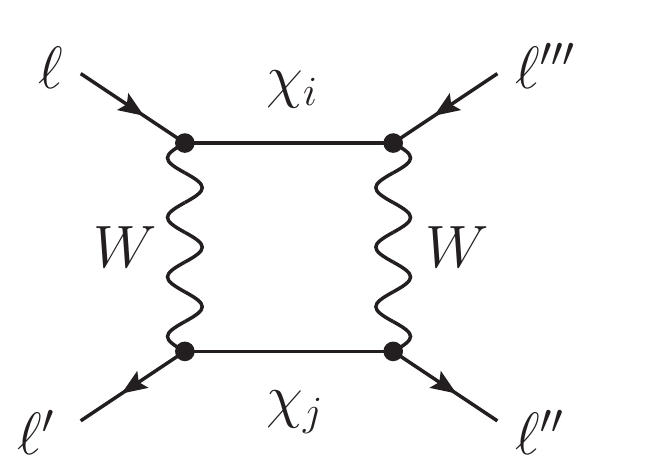} 
\end{center}
\caption{Box diagrams contributing to $\ell\to\ell'\ell''\bar{\ell}'''$. The diagram on the right introduces explicit LNV contributions.}
       \label{Box-3l-fig}          
\end{figure}

The box form factors are all finite. The amplitude for \decayS\ receives the contribution of diagrams with explicit lepton number violating (LNV) vertices (Fig.~\ref{Box-3l-fig}). To implement the LNV vertices we have followed the algorithm in \cite{Denner:1992vza} that circumvents the explicit introduction of the charge conjugation matrix in the Feynman rules and allows to use Dirac propagators also for Majorana particles. In particular, the diagrams on the right of Fig.~\ref{Box-3l-fig} contain genuine LNV contributions from Majorana particles that should vanish if lepton number is conserved. We have verified that this is indeed the case when the two fermion singlets form a Dirac field, {\it i.e.} when $\mu=0$ ($r=1$). The Lorentz structure of all box diagrams can be reduced to the form in (\ref{Box-contributions}) after some algebra (see Appendix~\ref{identities}). In agreement with \cite{Ilakovac:1994kj}, we find 
\eq{
F_B&=\frac{\alpha_W}{16\pi M_W^2 s_W^2}
\nonumber\\
&\times\Big\{
\sum_{i,j}^{5} \left[B^{*}_{\ell i} B_{\ell^{\prime}i} B^{*}_{\ell^{\prime\prime\prime}j} B_{\ell^{\prime\prime}j} + (\ell'\leftrightarrow\ell'') \right] f_{B_d}(x_i,x_j)
\nonumber\\
&\quad+\sum_{i,j}^{5} B^{*}_{\ell i} B_{\ell^{\prime}j} B^{*}_{\ell^{\prime\prime\prime}i} B_{\ell^{\prime\prime}j} f_{B}^{\rm LNV}(x_i,x_j)\Big\},\label{BoxII}
}
where
\eq{
f_{B_d}(x,y)&=
\left( 1+\frac{x y}{4}\right)\tilde{d}(x,y)-2xy\,d(x,y),\\
f^{\rm LNV}_{B}(x,y)&=\sqrt{xy}\left[2\tilde{d} (x,y)-(4+x y)d(x,y) \right],
}
and
\eq{
\tilde{d}(x,y)&=
 \frac{x^2\ln x}{(1-x)^2(y-x)}
+\frac{y^2\ln y}{(1-y)^2(x-y)}
\nonumber\\&
-\frac{1}{(1-x)(1-y)},
\\
d(x,y)&=
 \frac{x\ln x}{(1-x)^2(y-x)}
+\frac{y\ln y}{(1-y)^2(x-y)}
\nonumber\\&
-\frac{1}{(1-x)(1-y)}.
}

\begin{figure}
\begin{center}
   \includegraphics[scale=0.55]{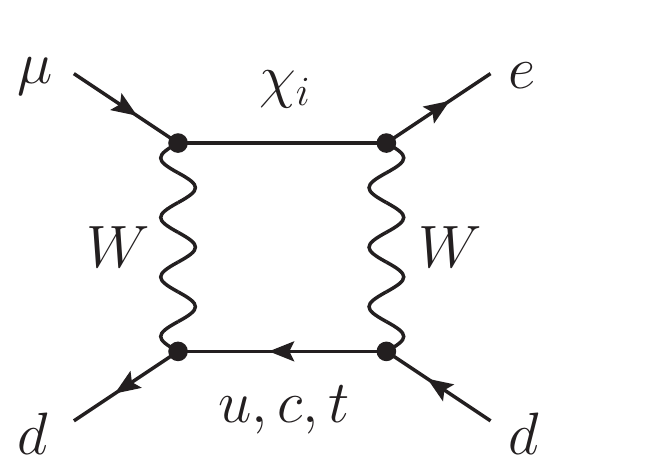} \quad 
   \includegraphics[scale=0.55]{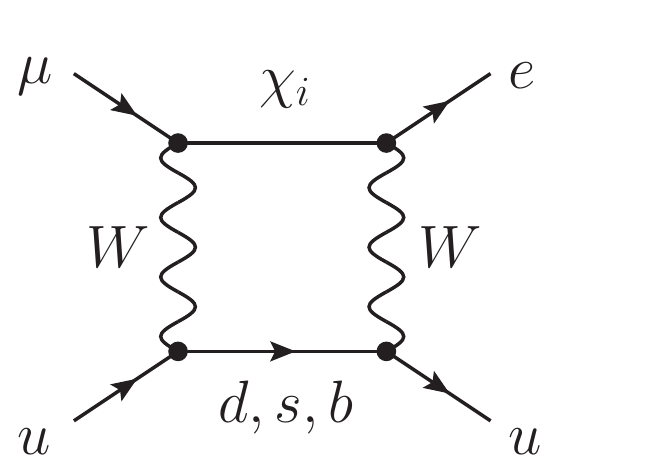} 
\end{center}
\caption{Box diagrams contributing to $\mu-e$ conversion in nuclei.}
       \label{mu-e-box-fig}          
\end{figure}

The one-loop contributions to the box form factors of $\mu-e$ conversion come from the diagrams in Fig.~\ref{mu-e-box-fig}. We obtain:
\eq{
F_{B_d}&=\frac{\alpha_W}{16\pi m_{W}^{2}s_{W}^{2}}\sum_{i}^{5}\sum_{j}^{3}B^{*}_{\mu i}B_{ei}\left|V_{jd}\right|^{2}f_{B_{d}}(x_{i}, x_{j}^{u}),
\\
F_{B_u}&=\frac{\alpha_W}{16\pi m_{W}^{2}s_{W}^{2}}\sum_{i}^{5}\sum_{j}^{3}B^{*}_{\mu i}B_{ei}\left|V_{uj}\right|^{2}f_{B_{u}}(x_{i}, x_{j}^{d}),
}
where $x^q_i=m^2_{q_i}/M_W^2$, $V_{ij}$ is the CKM matrix and
\eq{
f_{B_{u}}(x,y)&=-\left(4+\frac{x y}{4} \right)\tilde{d}(x,y)+2xy\,d(x,y).
}
Neglecting all quark masses, except that of the top quark, and defining $x_t=m_{t}^{2}/m_{W}^{2}$, we may write:
\eq{
\sum_j^3&\left|V_{jd}\right|^2 f_{B_{d}}(x_{i},x_{j}^{u})
\nonumber\\
&= \left|V_{td}\right|^2\left[f_{B_d}(x_i,x_t)-f_{B_{d}}(x_i,0)\right]
   -f_{B_{d}}(x_{i},0),\\
\sum_j^3&\left|V_{uj}\right|^2 f_{B_u}(x_i,x_j^d)= f_{B_{u}}(x_i,0).
}

In Appendix~\ref{onlyN} we show how to express all these form factors in terms of the contributions of heavy neutrinos only.


\section{Numerical results}
\label{NA}

Next we analyze the predictions of our model for different values of its free parameters: the three heavy-light mixings $s_{\nu_e}$, $s_{\nu_\mu}$ and $s_{\nu_\tau}$, the mass $m_{N_1}$ of the lightest heavy neutrino and the mass ratio $r=m_{N_2}^2/m_{N_1}^2$. If $r=1$ the two heavy Majorana neutrinos become
a single Dirac field. 

In order to be consistent with perturbative unitarity, the Yukawa couplings cannot exceed an upper limit. We will take
\eq{
 Y_{\nu_i}
=\frac{\sqrt{2 m_{N_1}m_{N_2}}}{v}s_{\nu_i}
=\frac{\sqrt{2}}{v}m_{N_1} r^{1/4} s_{\nu_i}
<\sqrt{4\pi}.
\label{PU-limit}
}
This means that, given $m_{N_1}$ and $r$, 
\eq{
s_{\nu_i}
< \frac{\sqrt{2\pi}v}{m_{N_1}r^{1/4}},
}
that constrains the mixings if $m_{N_1}r^{1/4}\gtrsim 620$~GeV. In particular, $s_{\nu_i}<0.12$ for $m_{N_1} r^{1/4}=5$~TeV. Given the mixings $s_{\nu_i}$ this condition also implies
\eq{
m_{N_1} r^{1/4}
< \frac{\sqrt{2\pi}v}{\max\{s_{\nu_i}\}} .
\label{PTU}
}

On the other hand, the heavy-light mixings must respect indirect constraints. 
We take $2\sigma$ limits from the global fit to electroweak precision observables and 
lepton flavor conserving processes in \cite{Fernandez-Martinez:2016lgt}, 
where the effects of extra neutrinos are encoded in effective operators:\footnote{%
In a recent work \cite{Coutinho:2019aiy} a global fit to modified neutrino couplings has been performed that alleviates the Cabibbo-angle anomaly and is compatible with the bounds we use.
}
\eq{
s_{\nu_e}<0.050,\quad 
s_{\nu_{\mu}}<0.021,\quad 
s_{\nu_{\tau}}<0.075. 
\label{ind-limits}
}
Then (\ref{PTU}) implies that $m_{N_1} r^{1/4}<8.2$~TeV if all mixings are fixed to the upper limits, but it could be larger otherwise.

\subsection{$\mu-e$ transitions}

\begin{figure}
    \includegraphics[scale=0.75]{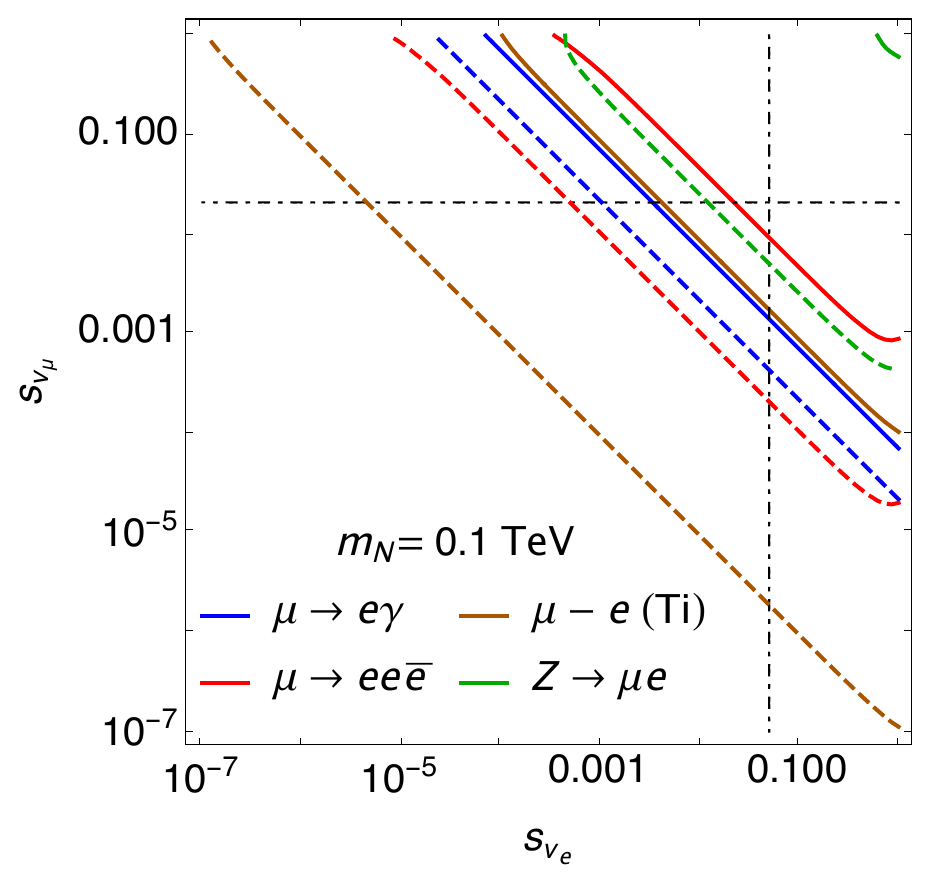}\quad
    \includegraphics[scale=0.75]{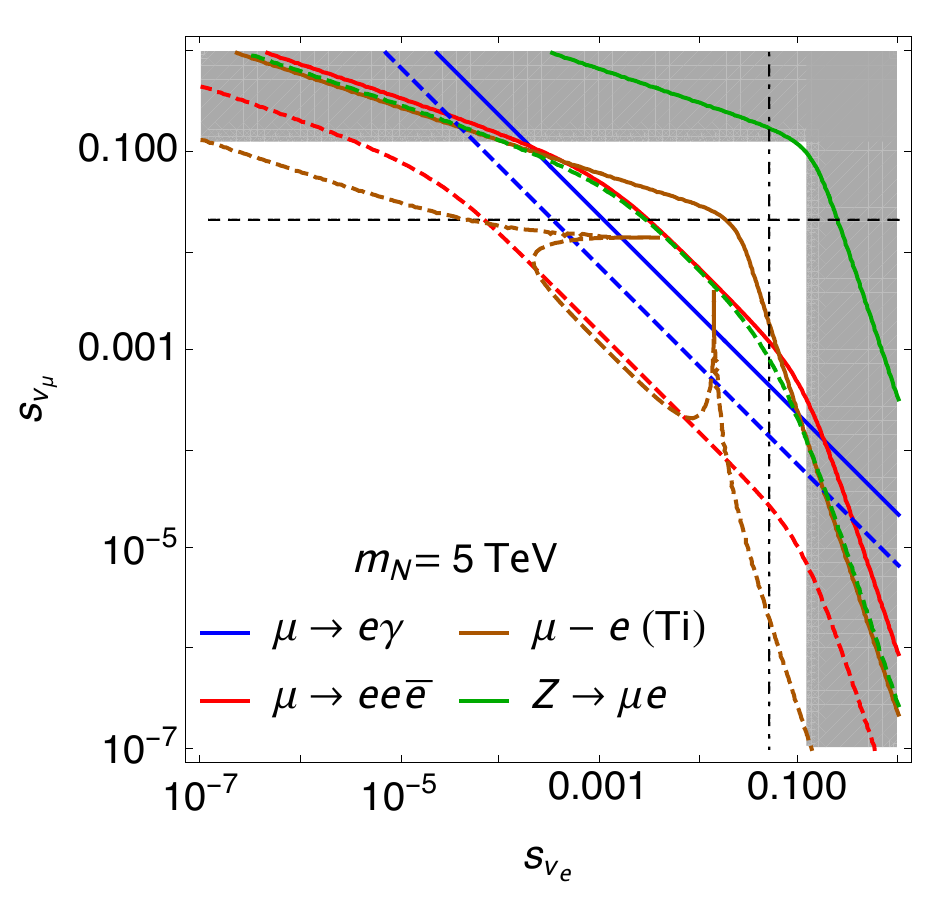}
    \caption{Contour plots in the $s_{\nu_e}-s_{\nu_\mu}$ plane assuming degenerate heavy neutrino masses $m_{N_1}=m_{N_2}=m_N=0.1,\,5$~TeV that saturate present limits (solid lines) and future sensitivities (dashed lines) of several $\mu-e$ transitions. A conservative value $s_{\nu_\tau}=0$ is assumed when needed. The black dot-dashed lines show current indirect limits. The gray-shadowed region is beyond the pertubative limit $Y_{\nu_i}^2<4\pi$.}   \label{Case21}  
\end{figure}

LFV processes involving only $\mu-e$ transitions further constrain the masses and mixings of the heavy neutrinos in our model. Let us first consider the case of one singlet Dirac neutrino ($m_{N_1}=m_{N_2}=m_N$). Figure~\ref{Case21} shows the contours in the $s_{\nu_e}-s_{\nu_{\mu}}$ plane that saturate present experimental bounds (solid lines) and the future sensitivities in Table~\ref{cLFV limits} (dashed lines) for $m_N=0.1,5$ TeV. In general the amplitudes for $Z\to\bar\mu e$, $\mu\to ee\bar{e}$ and $\mu-e$ conversion in nuclei depend on all three heavy-light mixing angles through the $Z$-penguin contribution, that involves the $C_{ij}$ matrix elements (\ref{Csfortwo}). Here we have assumed $s_{\nu_{\tau}}=0$, so the regions below the curves in Fig.~\ref{Case21} enclose the most conservative values for $s_{\nu_e}$ and $s_{\nu_\mu}$ ({\it i.e.}, a non-zero $s_{\nu_{\tau}}$ would imply stronger bounds). Actually, $\mu\to e \gamma$ sets stringent constrains only on the product $s_{\nu_e}s_{\nu_\mu}$; for $m_N\gtrsim 1$~TeV we find that this process does not depend on the heavy neutrino masses and that its branching fraction can be approximated by
\eq{
{\rm BR}(\mu \to e \gamma)\approx \frac{3 \alpha}{8 \pi}s_{\nu_e}^2s_{\nu_\mu}^2,
} 
which yields the conservative direct limit: 
\eq{
s_{\nu_e}^2s_{\nu_{\mu}}^2<5.1\times 10^{-10}. \label{limitmueg}
}
Fig.~\ref{Case21} also reveals that in forthcoming experiments $\mu-e$ (Ti) and $\mu\to ee\bar{e}$ will be more constraining than $\mu\to e\gamma$. 

Another point that we would like to emphasize is that the amplitudes for $\mu\to ee\bar{e}$, $\mu-e$ conversion in nuclei and $Z\to \mu e$ introduce terms of order $s_{\nu_{i}}^4$ that cannot be ignored, since they imply a strong quadratic dependence on the heavy neutrino masses. Indeed, these terms dominate the amplitude when the splitting between the two heavy masses is large. Our results differ then from those in \cite{Fernandez-Martinez:2016lgt}, where as a first approximation the terms proportional to $s_{\nu_i}^4$ are neglected.

From a phenomenological point of view it is also interesting to investigate whether the model can accommodate values of the different observables involving $\mu-e$ transitions near the current experimental bounds. What are the maximum values of $\mu\to e\gamma$, $\mu\to ee\bar{e}$, $\mu-e$ (Ti) and $Z\to \mu e$ consistent with all the bounds in heavy neutrino models? To answer this question we have considered the cases with low ($r=1$) and high ($r=25$) neutrino mass ratio and different values of $s_{\nu_e}s_{\nu_\mu}$, $m_N$, and $s_{\nu_\tau}$ that respect the indirect and perturbative limits.

\begin{figure}
    \includegraphics[scale=0.75]{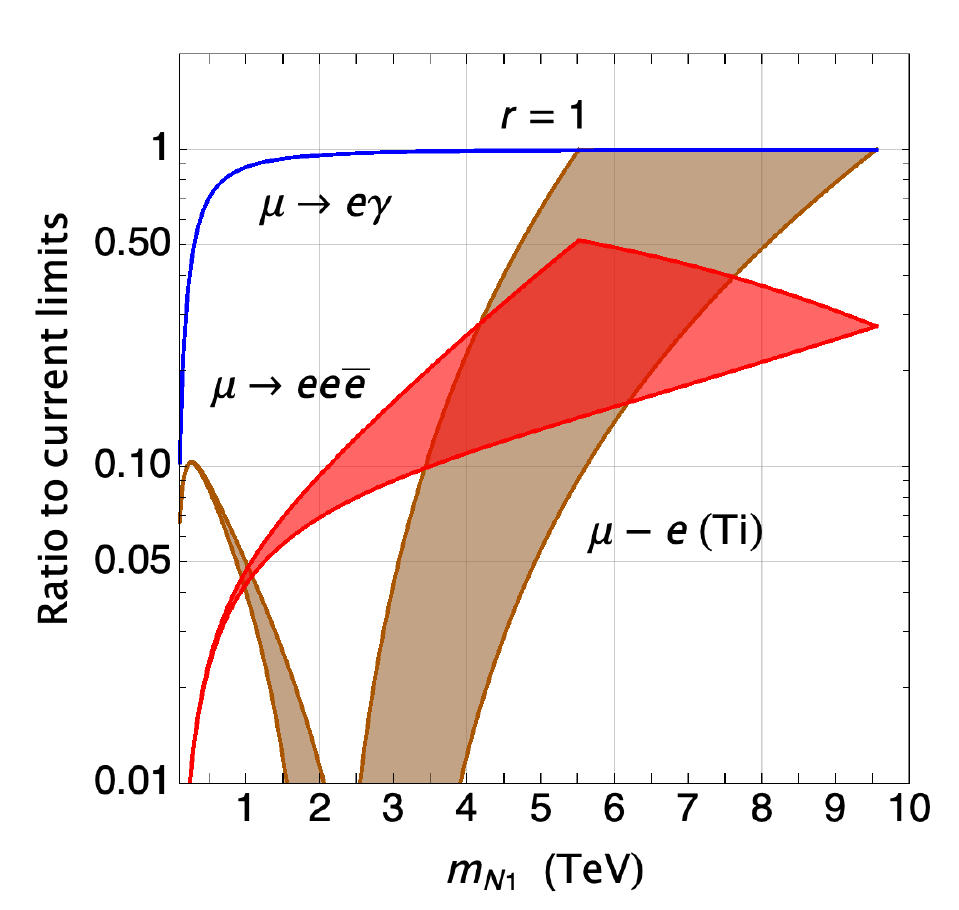} \quad
    \includegraphics[scale=0.75]{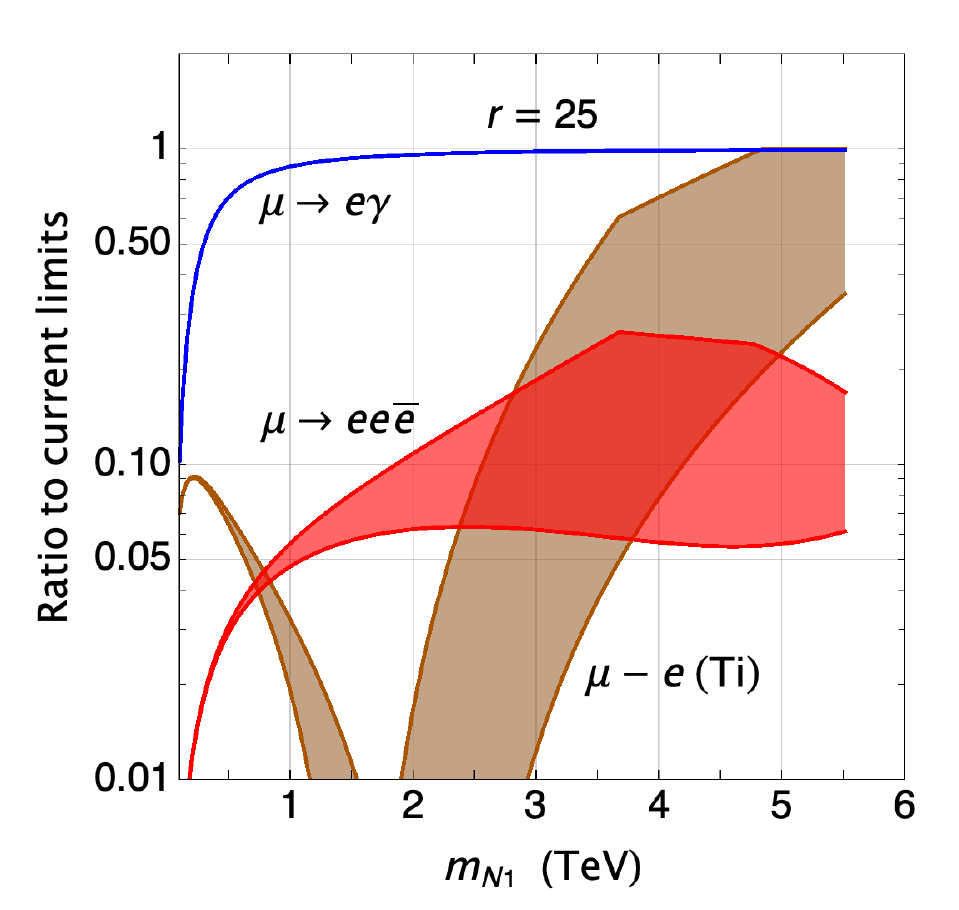}
\caption{%
Predictions normalized to current limits assuming fixed maximum mixings $s_{\nu_e}^{\rm max}$ and $s_{\nu_\mu}^{\rm max}$ compatible with $\mu\to e\gamma$ (\ref{limitmueg}) and the indirect limits (\ref{ind-limits}), for two values of the neutrino mass-ratio $r$. The predictions for $\mu\to ee\bar{e}$ and $\mu-e$~(Ti) get constrained to the corresponding bands whose upper (lower) boundaries are determined by $s_{\nu_\tau}=0\;(s_{\nu_\tau}^{\rm max})$. Higher masses are forbidden by perturbative unitarity (\ref{PTU}) for these mixings. The ratio of BR$(Z\to\mu e)$ to current limits is always much smaller than the others, below $10^{-6}$. 
}   
\label{Plotmue} 
\end{figure}

In Fig.~\ref{Plotmue} we plot the ratio of the different observables to their current bound. In general, it is $\mu\to e\gamma$ the most constraining process, so in the plot we set the maximum value of $s_{\nu_e}s_{\nu_\mu}$ compatible with that process and vary the rest of parameters. The lower (upper) curves of each band correspond to $s_{\nu_\tau}=0\;(s_{\nu_\tau^{\rm max}}=0.075)$, whereas the drop in the $\mu-e$ conversion amplitude at neutrino masses $\lesssim 2$~TeV is due to the opposite sign in the form factors $F_{B_d}$ and $F_{B_u}$.

We find that $\mu-e$ (Ti) may also saturate its present bounds if the neutrino masses are large enough: $m_N>5.5$ TeV for a Dirac neutrino ($r=1$) and $m_N>4.7$ TeV if $r=25$. In contrast, in these models the processes $\mu\to ee\bar{e}$ and $Z\to \mu e$ can not reach their current experimental limits consistently with $\mu\to e\gamma$ and $\mu-e$ (Ti)  for any values of the free parameters. We find 
\eq{
{\rm BR}(\mu \to e e \bar e)<5.2\times 10^{-13}, \label{b1}
}
and
\eq{
{\rm BR}(Z \to \mu e)<6.5\times 10^{-13}. \label{b2}
}
These limits do not change for lower values of $s_{\nu_e}s_{\nu_\mu}$, so our result implies that the observation at future experiments of any of these processes at a rate between the current bounds and these upper limits would exclude heavy neutrinos (both Dirac or Majorana) as a possible explanation. Larger values of the Majorana mass ratio $r$ than the one shown in the second plot of Fig.~\ref{Plotmue} would result in similar allowed regions just cutting off the higher masses to meet the perturbative unitarity limit. 

Looking at the improvement factor of the sensitivities in future experiments (Table~\ref{cLFV limits}) we conclude that $\mu-e$~(Ti) will take the lead in constraining the parameter space of our model, rather than $\mu\to e\gamma$, except for a tiny region of masses between 1.5 and 2~TeV that would be probed better by $\mu\to ee\bar{e}$.

\subsection{$\tau-e$ transitions}

\begin{figure}
    \includegraphics[scale=0.75]{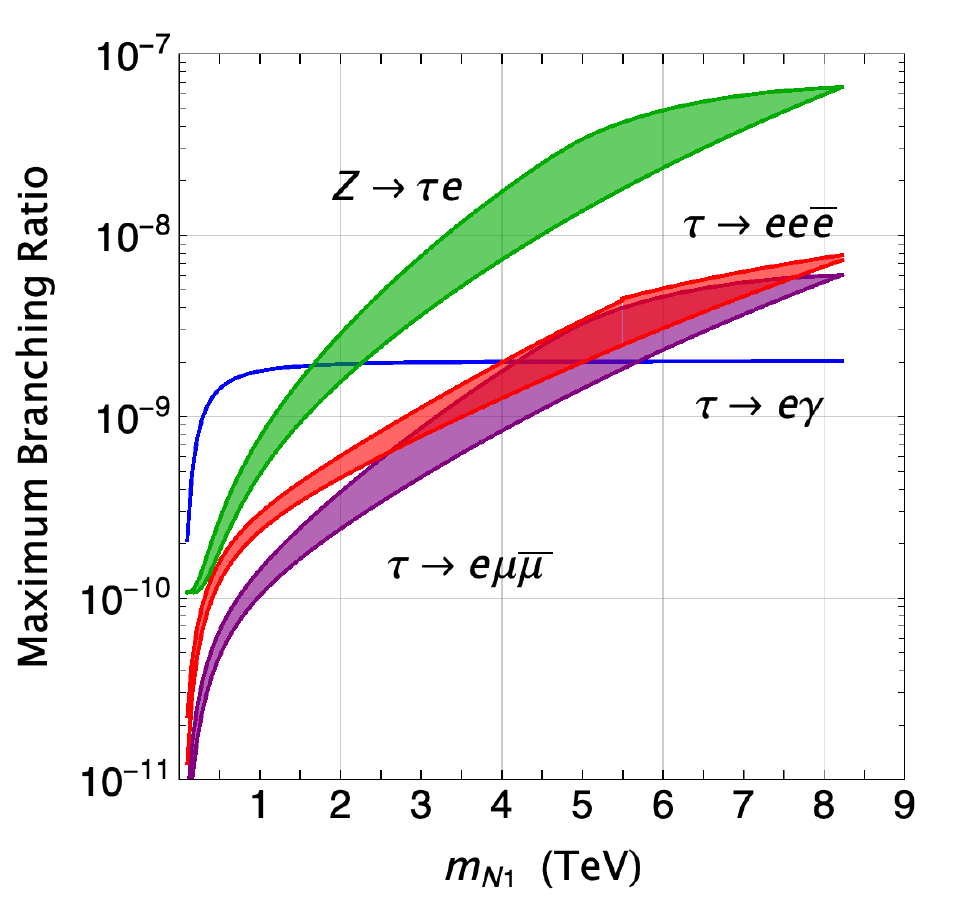} \quad
    \includegraphics[scale=0.75]{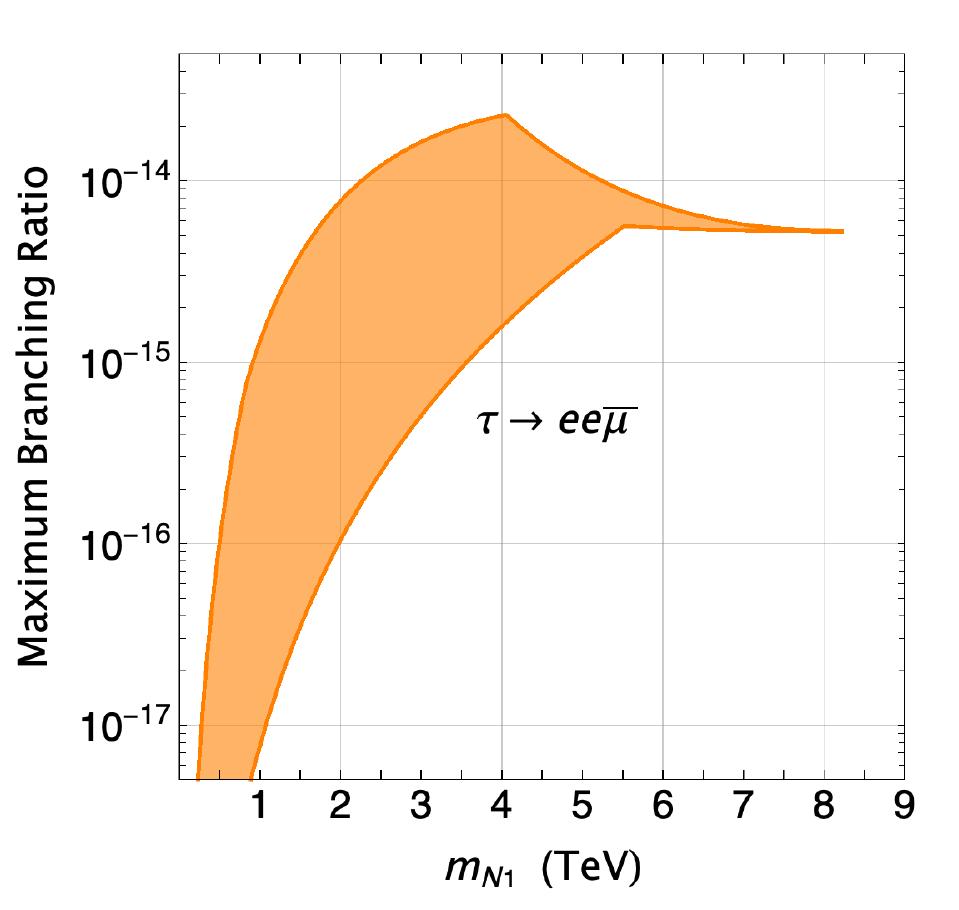}
\caption{Maximum values of $\tau-e$ transition rates compatible with current constraints on $\mu\to e\gamma$, $\mu-e$~(Ti) and the indirect limits (\ref{ind-limits}). These maximum rates get constrained to the corresponding bands whose lower (upper) boundaries are determined by $r=1\;(r\gg 1)$.} \label{Plottaue} 
\end{figure}

The constraints on our model from current limits on LFV $\tau-e$ transitions (Table~\ref{cLFV limits}) turn out to be less restrictive than those involving the first two lepton families. In Fig.~\ref{Plottaue} we show our predictions for the maximum possible rates for several LFV $\tau$ decays consistent with current bounds from $\mu\to e\gamma$ and $\mu-e$~(Ti), the indirect limits (\ref{ind-limits}) and perturbative unitarity (\ref{PTU}). 

The maximum branching ratio for $\tau\to e\gamma$, independent of heavy neutrino masses as for $\mu\to e\gamma$, is
\eq{
{\rm BR}(\tau\to e \gamma) < 2.0\times 10^{-9},
}
close but still below the future sensitivity of Belle-II.

The predictions for $\tau\to ee\bar{e}$ and $\tau\to e\mu\bar{\mu}$ are very similar because the dominant contribution comes in both cases from the $Z$ penguin diagram. They can reach:
\eq{
{\rm BR}(\tau\to ee\bar{e})& < 7.3\times 10^{-9},\\ 
{\rm BR}(\tau\to e\mu\bar{\mu})& < 6.0\times 10^{-9}, 
}
which, unlike $\tau\to e\gamma$, are well within the expected sensitivity of Belle-II. These maximum values correspond to a Dirac neutino singlet ($r=1$) with a mass just below the perturbative limit for the largest posible mixings, $m_N=8.2$~TeV. For smaller masses there is room for $r>1$ that enhance the decay rates up to the upper part of the shaded band in Fig~\ref{Plottaue}.

The decay $\tau\to ee\bar{\mu}$ (double flavor change) is generated through box diagrams only, so its amplitude is always proportional to $s_{\nu_e}^2s_{\nu_\tau}s_{\nu_\mu}$ and hence more suppressed than the other channels (see lower plot of Fig~\ref{Plottaue}), with a maximum at
\eq{
{\rm BR}(\tau\to ee\bar{\mu}) < 2.3\times 10^{-14}.
}
Nevertheless, it is important to remark that this decay is sensitive to the genuine effects of Majorana neutrinos encoded in the LNV vertices of one of its box contributions. Unlike the other processes, the rate of this for two non-degenerate Majorana neutrinos can be enhanced by more than two orders of magnitude when compared to the case of a Dirac singlet. In fact, the maximum branching ratio above is obtained for $r\approx 16.9$ and $m_{N_1}\approx4.1$~TeV.

For $Z\to \tau e$ our model predicts
\eq{
{\rm BR}(Z\to \tau e) < 6.0\times 10^{-8},
}
which is at the reach of future circular colliders.

Concerning the maximum values for the rates of $\tau-\mu$ transitions, we get similar results as above by exchanging $e$ and $\mu$ and applying some approximate correction factors. In particular, taking the maximal possible mixings from the indirect limits (\ref{ind-limits}), the processes $\tau\to\mu\gamma$, $\tau\to\mu\mu\bar{\mu}$, $\tau\to\mu e\bar{e}$ and $Z\to\tau\mu$ are suppressed by a factor of $(s_{\nu_\mu}^{\rm max}/s_{\nu_e}^{\rm max})^2\approx 0.18$, whereas $\tau\to\mu\mu\bar{e}$ is enhanced by $s_{\nu_e}^{\rm max}/s_{\nu_\mu}^{\rm max}\approx 2.4$. 


\section{Conclusions}
\label{Conclusions}

The lepton sector of the SM is still poorly known. In particular, we do not know whether the observed neutrinos are Dirac or Majorana particles or if the sector includes additional fermion singlets (sterile neutrinos). These extra neutrinos, if any, would enhance cLFV processes that are otherwise very suppressed by the tiny masses of the observed neutrinos.

In this work we have introduced the simplest neutrino model that captures all the effects that are relevant to these processes: a minimal number of Majorana neutrino fields (three active and two sterile), two of them heavy and the rest massless, allowing unsuppressed heavy-light mixings and the possibility of LNV encoded in the mass splitting of the heavy states ($\mu=m_{N_2}-m_{N_1}=(\sqrt{r}-1)m_{N_1}$). Larger splittings, however, imply an increasing amount of
fine tuning required to cancel loop
corrections that deform the proposed pattern, which is only stable in the lepton number conserving
case with $\mu=0$. In any case,
only five parameters describe the model, that are expressed in terms of the two heavy masses and the three heavy-light mixings ($s_{\nu_k}^2$, $k=e,\mu,\tau$). The model could be perturbed to account for the extremely light neutrino masses ($m_{\nu_i}<1$~eV) and the observed PMNS mixings, but this would have no impact on cLFV.

We have explored the predictions of our model for the most relevant reactions involving one or two flavor changes. We have presented analytical expressions for all of them and calculated their expected rates compatible with present direct and indirect limits. Our computation is exact at one loop, including all orders in the heavy-light mixings, and the genuine Majorana effects have been singled out. This work completes and updates previous results and is in agreement with an effective field theory analysis of the most general seesaw extension of the SM \cite{Coy:2018bxr}. We conclude that forthcoming LFV experiments will probe a significant fraction of the parameter space of models with heavy Majorana neutrinos.


\section*{Acknowledgments}

We would like to thank F. del \'Aguila and Jos\'e Santiago for helpful discussions.  
This work was supported in part by the Spanish Ministry of Science, Innovation and Universities, under grants FPA2016-78220-C3-1,2,3-P (fondos FEDER), and Junta de Andaluc{\'\i}a, grants FQM~101 and SOMM17/6104/UGR.
G.H.T. wants to acknowledge financial support from Conacyt through the program ``Estancia Postdoctoral en el Extranjero". 
The work of P.R. has been partially funded by Conacyt through the project 250628
(Ciencia B\'asica) and Fondo SEP-Cinvestav 2018 (project number 142).

\appendix


\section{Useful identities\label{identities}}

In the limit of zero external momenta the Lorentz structure of all box diagrams can be reduced to the same form using several identities based on the decomposition in the chiral basis of Dirac matrices, transpositions and Fierz rearrangements:
\eq{
    \bar{u}(p_1)\gamma^\mu\gamma^\alpha&\gamma^\nu P_L u(p)\;
    \bar{u}(p_2)\gamma_\nu\gamma_\alpha\gamma_\mu P_L v(p_3) \nonumber\\
&= 4\,\bar{u}(p_1)\gamma^\mu P_L u(p)\;
    \bar{u}(p_2)\gamma_\mu P_L v(p_3), \\
    \bar{u}(p_1)\gamma^\mu\gamma^\alpha&\gamma^\nu P_L u(p)\;
    \bar{u}(p_2)\gamma_\mu\gamma_\alpha\gamma_\nu P_L v(p_3) \nonumber\\
&=16\,\bar{u}(p_1)\gamma^\mu P_L u(p)\;
    \bar{u}(p_2)\gamma_\mu P_L v(p_3), \\
    \bar{u}(p_3)\gamma^\mu\gamma^\nu&P_L u(p)\;
    \bar{u}(p_2)\gamma_\mu\gamma_\nu P_L v(p_1) \nonumber\\
&=\;\;\;4\,\bar{u}(p_3)P_L u(p)\;
      \bar{u}(p_2)P_R v(p_1) \nonumber\\
&=-4\,\bar{v}(p)P_L v(p_3)\;
      \bar{u}(p_2)P_R v(p_1) \nonumber\\
&=-\tfrac{1}{2}\,\bar{v}(p)\gamma^\mu P_R v(p_1)\;
      \bar{u}(p_2)\gamma_\mu P_L v(p_3) \nonumber\\
&=-\tfrac{1}{2}\,\bar{u}(p_1)\gamma^\mu P_L u(p)\;
      \bar{u}(p_2)\gamma_\mu P_L v(p_3), \\
   \bar{u}(p_2)\gamma^\mu P_L & u(p)\;
   \bar{u}(p_1)\gamma_\mu P_L v(p_3) \nonumber\\
&=-\bar{u}(p_1)\gamma^\mu P_L u(p)\;
   \bar{u}(p_2)\gamma_\mu P_L v(p_3) .
\label{A4}
}


\section{Partial decay widths for LFV three-body decays}\label{3lep}

Given the generic form factors in Eqs.~(\ref{ampli3lep}), (\ref{photon-contributions}), (\ref{Z-contributions}) and (\ref{Box-contributions}) the expressions for the partial widths of the three types of decays in Table~\ref{channels} are \cite{delAguila:2019htj}:
\eq{
\Gamma_1&= \frac{\alpha^2 m_{\ell}^5}{96 \pi} 
\Big\{ 3 \left|A_{1L}\right|^2 +2 \left|A_{2R}\right|^2 \left(8\ln\frac{m_\ell}{m_{\ell^{\prime\prime}}}-13\right) 
\nonumber \\
&+2\left|F_{LL}\right|^2+\left|F_{LR}\right|^2+\tfrac{1}{2}\left|F_B\right|^2
-\big[6 A_{1L}A_{2R}^* - F_{LL}F_{B}^*
\nonumber \\
& - (A_{1L}-2A_{2R})(2F_{LL}^*+F_{LR}^* + F_{B}^*) 
 + {\rm h.c.}\big] \Big\},
\\
\Gamma_2&= \frac{\alpha^2 m_{\ell}^5}{96 \pi} 
\Big\{ 2 \left|A_{1L}\right|^2 +4 \left|A_{2R}\right|^2 \left(4\ln\frac{m_\ell}{m_{\ell^{\prime\prime}}}-7\right) 
\nonumber \\
&+\left|F_{LL}\right|^2+\left|F_{LR}\right|^2+\left|F_B\right|^2
-\big[4 A_{1L}A_{2R}^* - \tfrac{1}{2} F_{LL}F_{B}^*
\nonumber \\
& - (A_{1L}-2A_{2R})(F_{LL}^*+F_{LR}^* + \tfrac{1}{2}F_{B}^*) 
 + {\rm h.c.}\big] \Big\},
\\
\Gamma_3&=\frac{\alpha^2m_{\ell}^5 }{192 \pi}\left| F_B\right| {}^2,
}
where
\eq{
F_{LL}=-\frac{F_L^Z(0)g_L^Z }{M_Z^2}, \quad F_{LR}=-\frac{F_L^Z(0)g_R^Z }{M_Z^2}.
}
and the $Z$ couplings $g_{L,R}^Z$ are given in Eq.~(\ref{Zllcoup}).

\section{The $\mu-e$ conversion rate}\label{muecon}

In terms of the form factors in Eqs.~(\ref{ampli3lep}), (\ref{photon-contributions}), (\ref{Z-contributions}) and (\ref{Box-quarks}), the $\mu-e$ conversion rate in a nucleus with $Z$ protons and $N=A-Z$ neutrons is given by
\eq{
{\cal R}&=\frac{\alpha^5 Z_{\rm eff}^4}{\Gamma_{\rm Capt}Z}F_P^2 m_\mu^5
\big|2Z\left(A_{1L}+A_{2R}\right)
\nonumber\\
&-\left(2Z+N\right)(F_{LL}^u+F_{LR}^u+B_L^u)
\nonumber\\
&-\left(Z+2N\right)(F_{LL}^d+F_{LR}^d+B_L^d)\big|^2,
}
where
\eq{
F_{LL}^q&=-\frac{F_L^Z(0)g_{Lq}^Z}{M_Z^2}, \quad
F_{LR}^q=-\frac{F_L^Z(0)g_{Rq}^Z}{M_Z^2}, 
\\
g_{Lu}^{Z}&=\frac{1-\frac{4}{3}s_W^2}{2s_Wc_W}, \quad 
g_{Ru}^{Z}=-\frac{2s_W}{3c_W},
\\  
g_{Ld}^{Z}&=\frac{-1+\frac{2}{3}s_W^2}{2s_Wc_W}, \quad 
g_{Rd}^{Z}=\frac{s_W}{3c_W},
}
and the rest of parameters are in Table \ref{Para-Conv}.

\begin{table}
\caption{Input parameters for different nuclei \cite{Kitano:2002mt,Suzuki:1987jf}.\label{Para-Conv}}
\begin{ruledtabular}
\begin{tabular}{cccccc}
Nucleus & $N$  & $Z$  & $Z_{\rm eff}$ & $F_P$   & $\Gamma_{\rm capt}$ [GeV]        
\\ \hline
${27 \atop 13}$Al   &  14 & 13 & 11.5 & 0.64 & $4.6 \times 10^{-19}$\\
${48 \atop 22}$Ti   &  26 & 22 & 17.6 & 0.54 & $1.7 \times 10^{-18}$\\
${197 \atop 79}$Au  & 118 & 79 & 33.5 & 0.16 & $8.6 \times 10^{-18}$
\end{tabular}
\end{ruledtabular}
\end{table}


\section{Expressions in terms of massive neutrinos only}\label{onlyN}

Using the relations between mixings matrices (\ref{identities1}) and (\ref{identities2}) with $\ell\ne\ell'$, one may write the generic contributions to all form factors in terms of massive neutrinos only, $m_{N_1}=m_{\chi_4}$, $m_{N_2}=m_{\chi_5}$: 
\eq{
\sum_{i}^{5}B_{\ell i}^{*}B_{\ell' i}f(x_i)&=\sum_{i}^{2}B_{\ell N_i}^{*}B_{\ell' N_i}\left[f(x_{N_i})-f(0)\right],
}
\eq{
\sum_{i,j}^{5}&B_{\ell i}^{*}B_{\ell' j}C_{ij}^{*}f(x_i,x_j)
=\sum_{i,j}^{2}B_{\ell N_i}^{*}B_{\ell' N_j}
\nonumber\\
&\times
\big\{\delta_{N_i N_j}\left[f(x_{N_i},0)+f(0,x_{N_j})-2f(0,0)\right]
\nonumber\\
&+C_{N_i N_j}^{*}\big[f(x_{N_i}, x_{N_j})-f(x_{N_i},0)-f(0,x_{N_j})
\nonumber\\
&\qquad\qquad +f(0,0)\big]\big\},
}
\eq{
\sum_{i,j}^{5}&B_{\ell i}^{*}B_{\ell' i}B_{\ell'''j}^{*}B_{\ell''j}f(x_i,x_j)
\nonumber\\
&=\sum_{i,j}^{2}B_{\ell N_i}^{*}B_{\ell' N_j}\delta_{\ell'' \ell'''}\delta_{N_{i}N_{j}}\left[f(x_{N_i},0)-f(0,0)\right]\nonumber\\
&+\sum_{i,j}^{2}B_{\ell N_i}^{*}B_{\ell' N_i}B_{\ell''' N_j}^{*}B_{\ell'' N_j}\big[f(x_{N_i},x_{N_j})
\nonumber\\
&\qquad\qquad 
-f(x_{N_i},0)-f(0,x_{N_i})+f(0,0) \big].
}



\begin{thebibliography}{9}

\bibitem{Glashow:1961tr} 
  S.~L.~Glashow,
  Nucl.\ Phys.\  {\bf 22}, 579 (1961).

\bibitem{Weinberg:1967tq} 
  S.~Weinberg,
  Phys.\ Rev.\ Lett.\  {\bf 19}, 1264 (1967).
   
\bibitem{Salam:1968rm} 
  A.~Salam,
  Conf.\ Proc.\ C {\bf 680519}, 367 (1968).
  
\bibitem{Fukuda:1998mi} 
  Y.~Fukuda {\it et al.} [Super-Kamiokande Collaboration],
  Phys.\ Rev.\ Lett.\  {\bf 81}, 1562 (1998)
  [hep-ex/9807003].
  
\bibitem{Ahmad:2001an} 
  Q.~R.~Ahmad {\it et al.} [SNO Collaboration],
  Phys.\ Rev.\ Lett.\  {\bf 87}, 071301 (2001)
 [nucl-ex/0106015].
  
\bibitem{Ahmad:2002jz} 
  Q.~R.~Ahmad {\it et al.} [SNO Collaboration],
  Phys.\ Rev.\ Lett.\  {\bf 89}, 011301 (2002)
  [nucl-ex/0204008].

\bibitem{Mohapatra:1998rq} 
  R.~N.~Mohapatra and P.~B.~Pal,
  World Sci.\ Lect.\ Notes Phys.\  {\bf 60}, 1 (1998)
  [World Sci.\ Lect.\ Notes Phys.\  {\bf 72}, 1 (2004)].

\bibitem{Pontecorvo:1957qd} 
  B.~Pontecorvo,
  Sov.\ Phys.\ JETP {\bf 7}, 172 (1958)
  [Zh.\ Eksp.\ Teor.\ Fiz.\  {\bf 34}, 247 (1957)].

\bibitem{Maki:1962mu} 
  Z.~Maki, M.~Nakagawa and S.~Sakata,
  Prog.\ Theor.\ Phys.\  {\bf 28}, 870 (1962).

\bibitem{Cabibbo:1963yz} 
  N.~Cabibbo,
  Phys.\ Rev.\ Lett.\  {\bf 10}, 531 (1963).

\bibitem{Kobayashi:1973fv} 
  M.~Kobayashi and T.~Maskawa,
  Prog.\ Theor.\ Phys.\  {\bf 49}, 652 (1973).
        
\bibitem{Minkowski:1977sc} 
  P.~Minkowski,
  Phys.\ Lett.\  {\bf 67B}, 421 (1977).
  
\bibitem{GellMann:1980vs}
  M.~Gell-Mann, P.~Ramond and R.~Slansky,
  Conf.\ Proc.\ C {\bf 790927}  315 (1979)
  [arXiv:1306.4669 [hep-th]].
  
\bibitem{Mohapatra:1979ia} 
  R.~N.~Mohapatra and G.~Senjanovic,
  Phys.\ Rev.\ Lett.\  {\bf 44}, 912 (1980).

\bibitem{Petcov:1976ff} 
  S.~T.~Petcov,
  Sov.\ J.\ Nucl.\ Phys.\  {\bf 25}, 340 (1977)
  [Yad.\ Fiz.\  {\bf 25}, 641 (1977)]
  Erratum: [Sov.\ J.\ Nucl.\ Phys.\  {\bf 25}, 698 (1977)]
  Erratum: [Yad.\ Fiz.\  {\bf 25}, 1336 (1977)].

\bibitem{Bilenky:1977du}
  S.~M.~Bilenky, S.~T.~Petcov and B.~Pontecorvo,
  Phys.\ Lett.\  {\bf 67B}, 309 (1977).

    
\bibitem{Cheng:1985bj} 
  T.~P.~Cheng and L.~F.~Li,
  Oxford, Uk: Clarendon (1984) 536 P. (Oxford Science Publications).

\bibitem{Illana:1999ww} 
  J.~I.~Illana, M.~Jack and T.~Riemann,
  hep-ph/0001273.
  
\bibitem{Illana:2000ic} 
  J.~I.~Illana and T.~Riemann,
  Phys.\ Rev.\ D {\bf 63}, 053004 (2001)
 [hep-ph/0010193].

\bibitem{Hernandez-Tome:2018fbq} 
  G.~Hern\'andez-Tom\'e, G.~L\'opez Castro and P.~Roig,
  Eur.\ Phys.\ J.\ C {\bf 79}, no. 1, 84 (2019)
  [arXiv:1807.06050 [hep-ph]].

\bibitem{Blackstone:2019njl} 
  P.~Blackstone, M.~Fael and E.~Passemar,
  arXiv:1912.09862 [hep-ph].

\bibitem{Arganda:2004bz} 
  E.~Arganda, A.~M.~Curiel, M.~J.~Herrero and D.~Temes,
  Phys.\ Rev.\ D {\bf 71}, 035011 (2005)
  [hep-ph/0407302].
  
\bibitem{Dinh:2012bp}
  D.~N.~Dinh, A.~Ibarra, E.~Molinaro and S.~T.~Petcov,
  JHEP {\bf 1208} 125 (2012)
   Erratum: [JHEP {\bf 1309} 023 (2013)]
  [arXiv:1205.4671 [hep-ph]].

\bibitem{Dinh:2013vya}
  D.~N.~Dinh and S.~T.~Petcov,
  JHEP {\bf 1309}, 086 (2013)
  [arXiv:1308.4311 [hep-ph]].

\bibitem{Abada:2014cca} 
  A.~Abada, V.~De Romeri, S.~Monteil, J.~Orloff and A.~M.~Teixeira,
  JHEP {\bf 1504}, 051 (2015)
  [arXiv:1412.6322 [hep-ph]].

\bibitem{Arganda:2014dta} 
  E.~Arganda, M.~J.~Herrero, X.~Marcano and C.~Weiland,
  Phys.\ Rev.\ D {\bf 91}, no. 1, 015001 (2015)
  [arXiv:1405.4300 [hep-ph]].

\bibitem{DeRomeri:2016gum} 
  V.~De Romeri, M.~J.~Herrero, X.~Marcano and F.~Scarcella,
  Phys.\ Rev.\ D {\bf 95}, no. 7, 075028 (2017)
  [arXiv:1607.05257 [hep-ph]].

\bibitem{Lindner:2016bgg}
  M.~Lindner, M.~Platscher and F.~S.~Queiroz,
  Phys.\ Rept.\  {\bf 731}, 1 (2018)
  [arXiv:1610.06587 [hep-ph]].

\bibitem{Mohapatra:1986bd} 
  R.~N.~Mohapatra and J.~W.~F.~Valle,
  Phys.\ Rev.\ D {\bf 34}, 1642 (1986).
  
\bibitem{Bernabeu:1987gr} 
  J.~Bernab\'eu, A.~Santamar\'ia, J.~Vidal, A.~M\'endez and J.~W.~F.~Valle,
  Phys.\ Lett.\ B {\bf 187}, 303 (1987).
  
\bibitem{Malinsky:2005bi} 
  M.~Malinsky, J.~C.~Romao and J.~W.~F.~Valle,
  Phys.\ Rev.\ Lett.\  {\bf 95}, 161801 (2005)
  [hep-ph/0506296].
  
  \bibitem{delAguila:2005yi}
  F.~del Aguila, M.~Masip and J.~L.~Padilla,
  Phys.\ Lett.\ B {\bf 627}, 131 (2005)
  [hep-ph/0506063].

\bibitem{delAguila:2017ugt} 
  F.~del Aguila, L.~Ametller, J.~I.~Illana, J.~Santiago, P.~Talavera and R.~Vega-Morales,
  JHEP {\bf 1708}, 028 (2017)
  Erratum: [JHEP {\bf 1902}, 047 (2019)]
 [arXiv:1705.08827 [hep-ph]].

\bibitem{delAguila:2019mvp} 
  F.~del Aguila, J.~I.~Illana, J.~M.~Perez-Poyatos and J.~Santiago,
  JHEP {\bf 1912}, 154 (2019)
 [arXiv:1910.09569 [hep-ph]].

\bibitem{Abada:2014kba} 
  A.~Abada, M.~E.~Krauss, W.~Porod, F.~Staub, A.~Vicente and C.~Weiland,
  JHEP {\bf 1411}, 048 (2014)
  [arXiv:1408.0138 [hep-ph]].

\bibitem{Arganda:2015naa} 
  E.~Arganda, M.~J.~Herrero, X.~Marcano and C.~Weiland,
  Phys.\ Rev.\ D {\bf 93}, no. 5, 055010 (2016)
  [arXiv:1508.04623 [hep-ph]].

\bibitem{Arganda:2015ija} 
  E.~Arganda, M.~J.~Herrero, X.~Marcano and C.~Weiland,
  Phys.\ Lett.\ B {\bf 752}, 46 (2016)
  [arXiv:1508.05074 [hep-ph]].

\bibitem{ArkaniHamed:1998rs} 
  N.~Arkani-Hamed, S.~Dimopoulos and G.~R.~Dvali,
  Phys.\ Lett.\ B {\bf 429}, 263 (1998)
 [hep-ph/9803315].
  
\bibitem{Randall:1999ee} 
  L.~Randall and R.~Sundrum,
  Phys.\ Rev.\ Lett.\  {\bf 83}, 3370 (1999)
  [hep-ph/9905221].

\bibitem{Furry:1939qr} 
  W.~H.~Furry,
  Phys.\ Rev.\  {\bf 56}, 1184 (1939).

\bibitem{Zeldovich:1981da} 
  Y.~B.~Zeldovich and M.~Y.~Khlopov,
  Pisma Zh.\ Eksp.\ Teor.\ Fiz.\  {\bf 34}, 148 (1981).

\bibitem{Coito:2019wte} 
  L.~Coito, C.~Faubel and A.~Santamaria,
  arXiv:1912.10001 [hep-ph].

\bibitem{Ilakovac:1995km} 
  A.~Ilakovac, B.~A.~Kniehl and A.~Pilaftsis,
  Phys.\ Rev.\ D {\bf 52}, 3993 (1995)
  [hep-ph/9503456].
  
\bibitem{Ilakovac:1995wc} 
  A.~Ilakovac,
  Phys.\ Rev.\ D {\bf 54}, 5653 (1996)
  [hep-ph/9608218].
  
\bibitem{Gribanov:2001vv} 
  V.~Gribanov, S.~Kovalenko and I.~Schmidt,
  Nucl.\ Phys.\ B {\bf 607}, 355 (2001)
  [hep-ph/0102155].

\bibitem{Littenberg:1991rd} 
  L.~S.~Littenberg and R.~E.~Shrock,
  Phys.\ Rev.\ D {\bf 46}, R892 (1992).
  
\bibitem{Barbero:2002wm} 
  C.~Barbero, G.~L\'opez Castro and A.~Mariano,
  Phys.\ Lett.\ B {\bf 566}, 98 (2003)
  [nucl-th/0212083].

\bibitem{Amhis:2016xyh} 
  Y.~Amhis {\it et al.} [HFLAV Collaboration],
  Eur.\ Phys.\ J.\ C {\bf 77}, no. 12, 895 (2017)
  [arXiv:1612.07233 [hep-ex]].
  
\bibitem{Calibbi:2017uvl} 
  L.~Calibbi and G.~Signorelli,
  Riv.\ Nuovo Cim.\  {\bf 41},  1 (2018).

\bibitem{Adam:2013mnn} 
  J.~Adam {\it et al.} [MEG Collaboration],
  Phys.\ Rev.\ Lett.\  {\bf 110}, 201801 (2013)
 [arXiv:1303.0754 [hep-ex]].
  
\bibitem{Bellgardt:1987du} 
  U.~Bellgardt {\it et al.} [SINDRUM Collaboration],
  Nucl.\ Phys.\ B {\bf 299}, 1 (1988).
  
\bibitem{Bertl:2006up} 
  W.~H.~Bertl {\it et al.} [SINDRUM II Collaboration],
  Eur.\ Phys.\ J.\ C {\bf 47}, 337 (2006).
  
\bibitem{Aubert:2009ag} 
  B.~Aubert {\it et al.} [BaBar Collaboration],
  Phys.\ Rev.\ Lett.\  {\bf 104}, 021802 (2010)
 [arXiv:0908.2381 [hep-ex]].
  
\bibitem{Tanabashi:2018oca} 
  M.~Tanabashi {\it et al.} [Particle Data Group],
  Phys.\ Rev.\ D {\bf 98}, no. 3, 030001 (2018).
  
\bibitem{Nehrkorn:2017fyt} 
  A.~Nehrkorn [CMS Collaboration],
  Nucl.\ Part.\ Phys.\ Proc.\  {\bf 287-288}, 160 (2017).
    
\bibitem{Akers:1995gz} 
  R.~Akers {\it et al.} [OPAL Collaboration],
  Z.\ Phys.\ C {\bf 67}, 555 (1995).

\bibitem{Abreu:1996mj} 
  P.~Abreu {\it et al.} [DELPHI Collaboration],
  Z.\ Phys.\ C {\bf 73}, 243 (1997).
  
\bibitem{Khachatryan:2016rke} 
  V.~Khachatryan {\it et al.} [CMS Collaboration],
  Phys.\ Lett.\ B {\bf 763}, 472 (2016)
  [arXiv:1607.03561 [hep-ex]].
  
\bibitem{Sirunyan:2017xzt} 
  A.~M.~Sirunyan {\it et al.} [CMS Collaboration],
  JHEP {\bf 1806}, 001 (2018)
 [arXiv:1712.07173 [hep-ex]].
  
\bibitem{Baldini:2018nnn} 
  A.~M.~Baldini {\it et al.} [MEG II Collaboration],
  Eur.\ Phys.\ J.\ C {\bf 78}, no. 5, 380 (2018)
 [arXiv:1801.04688 [physics.ins-det]].
  
\bibitem{Blondel:2013ia} 
  A.~Blondel {\it et al.},
  arXiv:1301.6113 [physics.ins-det].

\bibitem{Alekou:2013eta} 
  A.~Alekou {\it et al.},
  arXiv:1310.0804 [physics.acc-ph].

\bibitem{Kou:2018nap} 
  E.~Kou {\it et al.} [Belle-II Collaboration],
  arXiv:1808.10567 [hep-ex]. To be published in PTEP.

\bibitem{Dam:2018rfz} 
  M.~Dam,
  SciPost Phys.\ Proc.\  {\bf 1}, 041 (2019)
 [arXiv:1811.09408 [hep-ex]].

\bibitem{Cerri:2018ypt}
  A.~Cerri {\it et al.},
  arXiv:1812.07638 [hep-ph].

\bibitem{Kuno:2013mha} 
  Y.~Kuno [COMET Collaboration],
  PTEP {\bf 2013}, 022C01 (2013).
  
 \bibitem{Hays:2017ekz} 
  C.~Hays, M.~Mitra, M.~Spannowsky and P.~Waite,
  JHEP {\bf 1705}, 014 (2017)
 [arXiv:1701.00870 [hep-ph]].
       
\bibitem{Aaij:2014azz} 
  R.~Aaij {\it et al.} [LHCb Collaboration],
  JHEP {\bf 1502}, 121 (2015)
 [arXiv:1409.8548 [hep-ex]].

\bibitem{Blondel:2019yqr} 
  A.~Blondel {\it et al.},
  arXiv:1906.02693 [hep-ph].

\bibitem{CEPCStudyGroup:2018ghi} 
  J.~B.~Guimaraes da Costa {\it et al.} [CEPC Study Group],
  arXiv:1811.10545 [hep-ex].
 

 \bibitem{Bolton:2019pcu}
  P.~D.~Bolton, F.~F.~Deppisch and P.~S.~B.~Dev,
  arXiv:1912.03058 [hep-ph].
 

\bibitem{Ilakovac:1994kj} 
  A.~Ilakovac and A.~Pilaftsis,
  Nucl.\ Phys.\ B {\bf 437}, 491 (1995)
  [hep-ph/9403398].
 
\bibitem{Hollik:1998vz} 
  W.~Hollik, J.~I.~Illana, S.~Rigolin, C.~Schappacher and D.~Stockinger,
  Nucl.\ Phys.\ B {\bf 551}, 3 (1999)
  Erratum: [Nucl.\ Phys.\ B {\bf 557}, 407 (1999)]
 [hep-ph/9812298].

\bibitem{Passarino:1978jh} 
  G.~Passarino and M.~J.~G.~Veltman,
  Nucl.\ Phys.\ B {\bf 160}, 151 (1979).
      
\bibitem{Hahn:1998yk} 
  T.~Hahn and M.~P\'erez-Victoria,
  Comput.\ Phys.\ Commun.\  {\bf 118}, 153 (1999)
 [hep-ph/9807565].
    
\bibitem{Denner:2016kdg} 
  A.~Denner, S.~Dittmaier and L.~Hofer,
  Comput.\ Phys.\ Commun.\  {\bf 212}, 220 (2017)
 [arXiv:1604.06792 [hep-ph]].

\bibitem{Patel:2016fam} 
  H.~H.~Patel,
  Comput.\ Phys.\ Commun.\  {\bf 218}, 66 (2017)
 [arXiv:1612.00009 [hep-ph]].

\bibitem{Denner:1992vza} 
  A.~Denner, H.~Eck, O.~Hahn and J.~Kublbeck,
  Nucl.\ Phys.\ B {\bf 387}, 467 (1992).
  
\bibitem{Fernandez-Martinez:2016lgt} 
  E.~Fern\'andez-Mart\'inez, J.~Hern\'andez-Garc\'ia and J.~L\'opez-Pavon,
  JHEP {\bf 1608}, 033 (2016)
  [arXiv:1605.08774 [hep-ph]].

\bibitem{Coutinho:2019aiy} 
  A.~M.~Coutinho, A.~Crivellin and C.~A.~Manzari,
  arXiv:1912.08823 [hep-ph].


\bibitem{Coy:2018bxr}
  R.~Coy and M.~Frigerio,
  Phys.\ Rev.\ D {\bf 99} (2019) no.9,  095040
  [arXiv:1812.03165 [hep-ph]].


\bibitem{delAguila:2019htj} 
  F.~del Aguila, L.~Ametller, J.~I.~Illana, J.~Santiago, P.~Talavera and R.~Vega-Morales,
  JHEP {\bf 1907}, 154 (2019)
 [arXiv:1901.07058 [hep-ph]].

\bibitem{Kitano:2002mt} 
  R.~Kitano, M.~Koike and Y.~Okada,
  Phys.\ Rev.\ D {\bf 66}, 096002 (2002)
  Erratum: [Phys.\ Rev.\ D {\bf 76}, 059902 (2007)]
  [hep-ph/0203110].

\bibitem{Suzuki:1987jf} 
  T.~Suzuki, D.~F.~Measday and J.~P.~Roalsvig,
  Phys.\ Rev.\ C {\bf 35}, 2212 (1987).

\end{thebibliography}
\end{document}